\RequirePackage{fix-cm}
\documentclass[smallextended]{svjour3}       % onecolumn (second format)
\smartqed  % flush right qed marks, e.g. at end of proof
\usepackage{graphicx}
\journalname{Journal of Ambient Intelligence and Humanized Computing}

\usepackage{url}
\usepackage{natbib}
\usepackage{subfigure}
\usepackage{amsmath}
\usepackage{amssymb}
\usepackage{threeparttable}
\usepackage{color}
\usepackage{multirow}
\usepackage[colorlinks,linkcolor=blue,citecolor=blue]{hyperref}
\usepackage[ruled,vlined]{algorithm2e}
\usepackage{bm}

\begin{document}
\newcommand{\tre}[1]{\textcolor{black}{#1}}
\newcommand{\tgr}[1]{\textcolor{black}{#1}}
\newcommand{\tbl}[1]{\textcolor{black}{#1}}

\title{DPN: Detail-Preserving Network with High Resolution Representation for Efficient Segmentation of Retinal Vessels\thanks{This work is supported by PhD research startup foundation of Xi'an University of Architecture and Technology (No.1960320048).}
}
%\subtitle{xx}

\titlerunning{DPN for Retinal Vessel Segmentation} % if too long for running head

\author{Song Guo}


\institute{Song Guo\at
              School of Information and Control Engineering, Xi'an University of Architecture and Technology,  Xi'an 710055, China \\
              \email{guomugong@hotmail.com}           %  \\
}

\date{Received: date / Accepted: date}

\maketitle

\begin{abstract}
Retinal vessels are important biomarkers for many ophthalmological and cardiovascular diseases. Hence, it is of great significance to develop automatic models for computer-aided diagnosis. Existing methods, such as U-Net follows the encoder-decoder pipeline, where detailed information is lost in the encoder in order to achieve a large field of view. Although spatial detailed information could be recovered partly in the decoder, while there is noise in the high-resolution feature maps of the encoder. And, we argue this encoder-decoder architecture is inefficient for vessel segmentation.
In this paper, we present the detail-preserving network (DPN), which avoids the encoder-decoder pipeline. To preserve detailed information and learn structural information simultaneously, we designed the detail-preserving block (DP-Block). Further, we stacked eight DP-Blocks together to form the DPN. More importantly, there are no down-sampling operations among these blocks. Therefore, the DPN could maintain a high/full resolution during processing, avoiding the loss of detailed information. To illustrate the effectiveness of DPN, we conducted experiments over three public datasets. Experimental results show, compared to state-of-the-art methods, DPN shows competitive/better performance in terms of segmentation accuracy, segmentation speed, and model size. Specifically,
1) Our method achieves comparable segmentation performance on the DRIVE, CHASE\_DB1, and HRF datasets.
2) The segmentation speed of DPN is over 20-160$\times$ faster than other methods on the DRIVE dataset.
3) The number of parameters of DPN is around 120k, far less than all comparison methods.
\end{abstract}

\section{Introduction}
\label{sec:intr}
Retinal blood vessels are an important part of fundus images,
and they can be applied to the diagnosis of many ophthalmological diseases, such as diabetic retinopathy~\citep{wong2018guidelines}, cataract~\citep{cao2020hierarchical}, and hypertensive
retinopathy~\citep{irshad2014classification}.
Specifically, when patients with diffuse choroidal hemangioma, retinal blood vessels will expand~\citep{scott1999diffuse}.
Vascular structures in patients with cataracts are unclear or even invisible~\citep{cao2020hierarchical}.
In addition, as retinal blood vessels and cerebral blood vessels are similar in
anatomical, physiological, and embryological characteristics, so that retinal vessels are also important biomarkers to some cardiovascular diseases~\citep{wong2004retinal,schmidt2018artificial}.
Accurate segmentation of blood vessels is the basic step of efficient computer-aided diagnosis (CAD).
However, manual segmentation of retinal vessels is time-consuming and relies heavily on the human experience.
Therefore, it is necessary to develop accurate and fast vessel segmentation methods for CAD.

%problems unsolved
Considering the clinical application scenarios, a good vessel segmentation model for CAD should satisfy the following two conditions.
1) High accuracy. The model needs to be capable to recognize both thin vessels and thick vessels, even for extremely thin vessels with one-pixel width. For example, the appearance of neovascularization can be used to diagnose and grade diabetes retinopathy~\citep{wong2018guidelines}.
2) Fast processing speed~\citep{yu2012fast,villalobos2010fast}.
The model needs to have a fast processing speed to meet clinical application, as faster speed means greater throughput and higher processing efficiency.

Existing vessel segmentation methods could be divided into two categories~\citep{GUO2019BTS}: unsupervised methods and supervised methods.
Unsupervised methods utilize manually designed low-level features and rules~\citep{azzopardi2013automatic,azzopardi2015trainable,srinidhi2018visual}, therefore, they show poor extensibility.
Supervised methods utilize human annotated training images, and their segmentation accuracy is usually higher than that of unsupervised methods~\citep{schmidt2018artificial}. Deep learning-based supervised methods could learn high-level features in an end-to-end manner, and they show superior performance in terms of segmentation accuracy and extensibility~\citep{jin2019dunet,threestage}.
Most deep vessel segmentation models follow the architecture of the fully convolutional network (FCN)~\citep{shelhamer2017fully}, in which the resolution of features is first down-sampled and then up-sampled to generate pixel-wise segmentation maps. However, the detailed information is lost in FCN.
Furthermore, a U-Net~\citep{ronneberger2015u} model was proposed, which could utilize intermediate layers in the up-sampling path to fuse more spatial information to generate fine segmentation maps.
Although the detailed information could be utilized in the U-Net, the extra noise was also introduced. Moreover, most U-Net variant models~\citep{jin2019dunet} require multiple forward passes to generate a segmentation map for one testing image, since they split one fundus image into hundreds of small patches.
As a result, they show slow segmentation speed and the contextual information is not fully utilized.

%our work
Different from U-Net that recovering the spatial details in the decoder to achieve a high-resolution representation, in this paper, we present a deep model termed detail-preserving network (DPN) which could preserve a high-resolution representation all the time.
Inspired by HRNet~\citep{wang2020deep}, the DPN learns the full-resolution representation directly rather than the low-resolution representation.
In this manner, the DPN could locate the boundaries of thin vessels accurately.
To this end, on one hand, we present the detail-preserving block (DP-Block), where multi-scale features are fused in a cascaded manner so that more contextual information could be utilized. And, the resolution of input features and output features of DP-Block is never changed, so that the detailed spatial information could be preserved.
On the other hand, we stacked eight DP-Blocks together to form the DPN. We note that there are no down-sampling operations among these DP-Blocks so that the DPN could learn both semantic features via a large field of view and preserve the detailed information simultaneously.
To validate the effectiveness of our method, we conducted experiments on the DRIVE, CHASE\_DB1, and HRF datasets. Experimental results reveal that our method shows competitive/better performance compared with other state-of-the-art methods.

Overall, our contributions are summarized as follows.
\begin{enumerate}
\item We present the detail-preserving block, which could learn the structural information and preserve the detailed information via intra-block multi-scale fusion.
\item We present the detail-preserving network, which mainly consists of eight serially connected DP-Blocks, and it maintains high-resolution representations during the whole process. As a result, the DPN could learn both semantic features and preserve the detailed information simultaneously.
\item We conducted experiments over three public datasets. Experimental results reveal that our method achieves comparable or even superior performance over other methods in terms of segmentation accuracy, segmentation speed, and model size.
\end{enumerate}

The rest of this paper is organized as follows. Related works about vessel segmentation are introduced in Section~\ref{sec:related}. Our method is described in Section~\ref{sec:dpn}. Experimental results are analyzed in Section~\ref{sec:exp}. Conclusions are drawn in Section~\ref{sec:con}.

\section{Related Works}
\label{sec:related}
Retinal vessel segmentation is a pixel-wise binary classification problem, and the objective is to locate each vessel pixel accurately for further processing. According to whether annotations are used, existing methods could be divided into two categories: unsupervised methods and supervised methods.

\subsection{Unsupervised Methods}
Unsupervised methods usually utilize human-designed low-level features, such as edge, line, and color. Manually annotated information is not utilized.
Unsupervised methods can be roughly divided into match filter based method~\citep{wang2013retinal,azzopardi2015trainable}, vessel tracking based method~\citep{yin2012retinal}, threshold based method~\citep{li2006multiscale,saleh2011automated} and morphology based method~\citep{garg2007unsupervised,wang2019retinal}.

Wang et al.~\citep{wang2013retinal} proposed a multi-stage method for vessel segmentation. In their method, a matched filtering was first adopted for vessel enhancing, and then vessels were located via a multi-scale hierarchical decomposition. Yin et al.~\citep{yin2012retinal} proposed a vessel tracking method, in which local grey information was utilized to select vessel edge points. Then a Bayesian method was used to determine the direction of vessels. Garg et al.~\citep{garg2007unsupervised} proposed a curvature-based method. In their method, the vessel lines were first extracted using curvature information, and then a region growing method was used to generate the whole vessel tree. Li et al.~\citep{li2006multiscale} proposed an adaptive threshold method for vessel segmentation, and their method could detect both large and small vessels. Christodoulidis et al.~\citep{christodoulidis2016multi} utilized line detector and tensor voting for vessel segmentation, and thin vessels were well detected.

A major limitation of the unsupervised method is that the features and rules are designed by a human. It is hard to design a satisfactory feature that works well on large-scale fundus images. This kind of method may show poor generalization ability.

\subsection{Supervised Methods}
In contrast to unsupervised methods, supervised methods need annotation information to build vessel segmentation models. Before deep learning methods were applied to vessel segmentation, supervised methods usually consist of two procedures: feature extraction and classification. In the first procedure, features were extracted by human-designed rules, just as that did in unsupervised methods. In the second procedure, supervised classifiers were employed to classify these extracted features into vessels or non-vessels. As deep learning methods unify feature extraction and classification procedures together, they could extract much discriminative features.

Deep learning-based methods could be roughly divided into classification-based methods and segmentation-based methods~\citep{srinidhi2017recent,mookiah2020review}. For classification-based methods, the category for each pixel is determined by its surrounding small image patch~\citep{Liskowski2016Segmenting,wang2019blood}.
This kind of method does not make full use of contextual information.
For segmentation-based methods, existing methods follow the architecture of FCN, where the resolution of feature maps are first down-sampled to encode structural information, and then the resolution of feature maps are up-sampled further to generate pixel-wise segmentation maps.
Although successive down-sampling operations could reduce the model's computational complexity and increase the model's receptive field, it inevitably loses detailed information.
As a result, this kind of method shows poor performance in the segmentation of thin/tiny blood vessels. To alleviate this problem, multi-scale fusion methods and graph models were adopted. For instance,
Maninis et al.~\citep{driu} proposed a FCN for vessel segmentation. They adopted a multi-scale feature fusion to generate fine vessel maps. Fu et al.~\citep{Fu2016DeepVessel} adopted a holistically-nested edge detection model~\citep{xie2017holistically} to generate coarse segmentation maps, and then a conditional random field was adopted to model the relationship among long-range pixels to refine segmentation maps.
Besides above methods, Ronneberger et al. proposed an u-shape network, called U-Net to preserve spatial information~\citep{ronneberger2015u}. Similar to FCN, the feature maps were first down-sampled to a low resolution, then they were up-sampled step-by-step.
In each step, the intermediate features with high representation in the encoder were utilized.
Several methods based on U-Net have been proposed for vessel segmentation. For instance, Jin et al.~\citep{jin2019dunet} proposed a DUNet for vessel segmentation. They used deformable convolution rather than grid convolution in U-Net to capture the shape of vessels. Wu et al.~\citep{wu2018} designed a two-branch network, where each branch consists of two U-Nets. The output of their method was the average of the predictions of these two branches.
In addition, different from~\citep{driu} and~\citep{Fu2016DeepVessel} that used the entire image as training samples.
Both~\citep{jin2019dunet} and~\citep{wu2018} used overlapped image patches of size 48$\times$48 as training samples, and a re-composed procedure is required to complete a segmentation map during testing.
Hence, they suffer from a high computation complexity.
Despite their success, the problem of losing spatial information in the down-sampling phase has not been fully addressed. Meantime, considering both computation complexity and segmentation accuracy, there still lacks a fast and accurate vessel segmentation model.

\section{Our Method}
\label{sec:dpn}

In this section, we will describe our method in detail, including the architecture of our proposed detail-preserving network,
the detail-preserving block, and the loss function at last.

\subsection{Detail-Preserving Network}
\begin{figure}
\centering
\subfigure[U-Net]{
\includegraphics[width=0.7\textwidth]{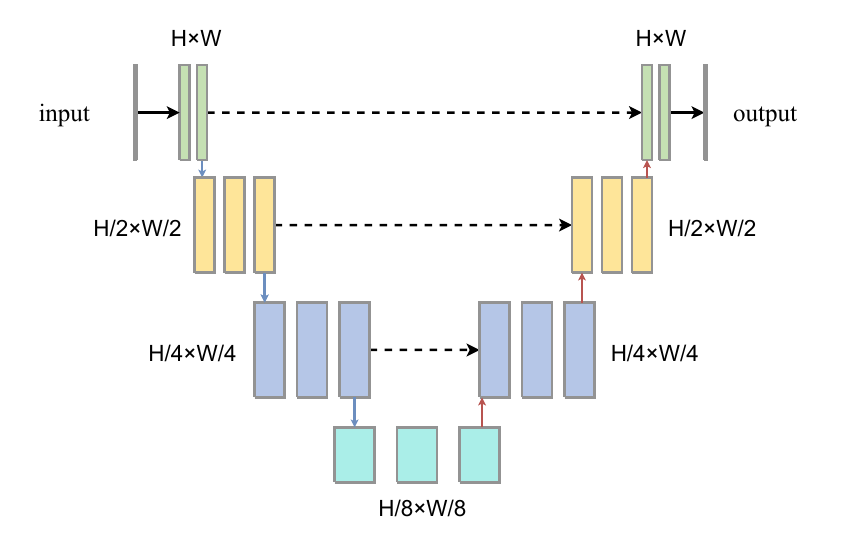}
\label{fig:unet_arch}
}
\subfigure[Our Method]{
\includegraphics[width=0.18\textwidth]{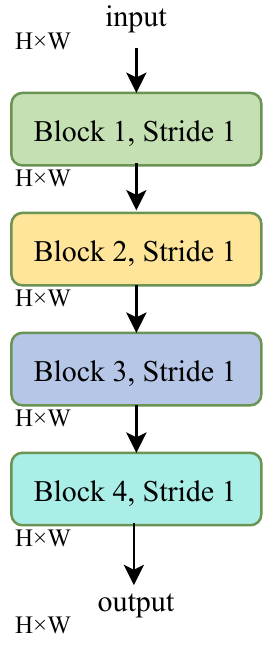}
}
\caption{(a) The architecture of U-Net~\citep{ronneberger2015u}. The resolution of feature maps is first decreased in the encoder, and then up-scaled in the decoder (H and W denote the height and width of feature maps). (b) The architecture of the proposed DPN, in which high-resolution representations are learned.}
\label{fig:unet_vs_dpn}
\end{figure}

A good vessel segmentation model should segment both thick vessels and thin vessels, this requires the segmentation model to learn structural semantic information and preserve detailed spatial information simultaneously.
The structural information is beneficial to locate thick vessels, and it requires the model to have a large field of view. While the detailed spatial information is important to locate vessel boundaries accurately, especially for thin vessels. However, it is easy to lose detailed information when learning structural information.
For example, the structural information of U-Net~\citep{ronneberger2015u} is learned by successive down-sampling operations, and the resolution of feature maps is decreased by a factor of 8 or even more (as can be seen in Fig.~\ref{fig:unet_arch}). Such low resolution implies that the spatial information of thin vessels is lost.
U-Net utilizes intermediate features of the encoder to recover the spatial information. However, intermediate feature maps of the encoder may have noise (non-vessel pixels are highlighted) due to a small field of view.

\begin{figure}[!t]
\centering
\subfigure{
\includegraphics[width=0.90\textwidth]{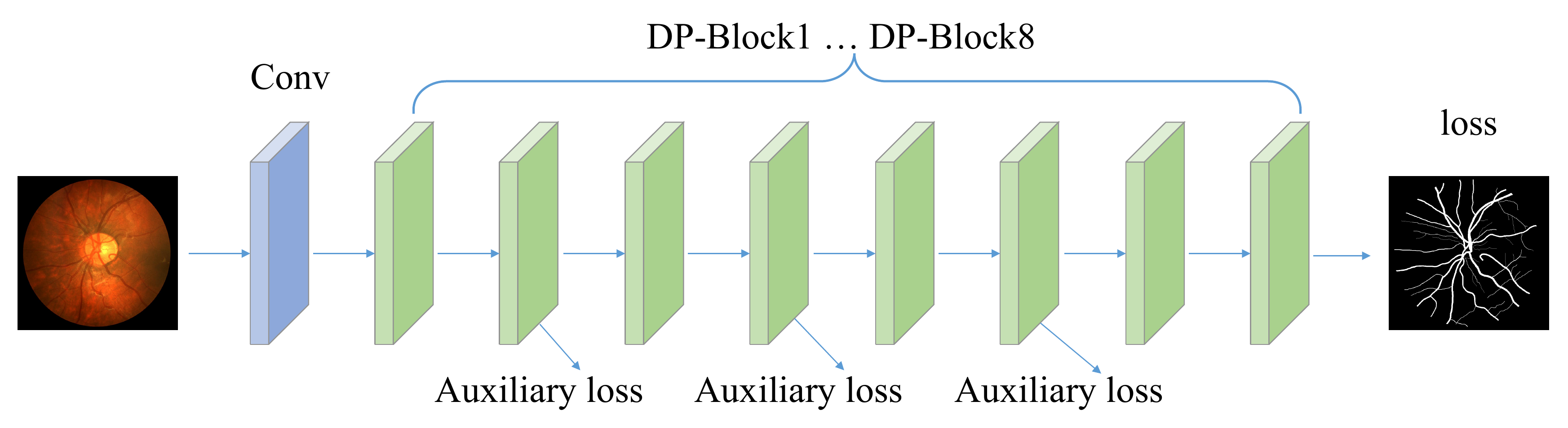}
}
\caption{Overview of the proposed detail-preserving network. DPN consists of one convolutional layer and eight DP-Blocks, and it maintains high/full resolution representations during the whole process.
Meantime, we add three auxiliary losses to pass extra gradient signals.}
\label{fig:net_overview}
\end{figure}

Our study is motivated by whether it is possible to preserve detailed information, while the network has a large field of view.
To this end, we present a high representation network, called detail-preserving network for vessel segmentation.
The architecture of our model is visualized in Fig.~\ref{fig:net_overview}.

We can observe from Fig.~\ref{fig:net_overview} that DPN mainly consists of three parts: the front convolution operation, eight detail-preserving blocks, and four loss functions (including three supervision losses). Compared with other vessel segmentation models, the DPN has the following characteristics.
1) Different from U-Net, there are no down-sampling operations among these DP-Blocks, this implies the resolution of features among these DP-Blocks keeps the same.
In other words, the DPN maintains a full-resolution representation during the whole processing (from input to output), thereby it could preserve detailed spatial information.
2) For DP-Block, the receptive field of the output neuron could be as large as four times that of the input neuron, while the detailed information could also be preserved.
Therefore, DPN could obtain a large field of view through stacking multiple DP-Blocks. Moreover, a large field of view means that DPN could learn semantic information instead of local information.
The architecture of the DP-Block will be described in the next section.
3) Different from U-Net that utilized VGGNet or ResNet as the backbone, which incurs a large number of parameters. The total number of parameters of DPN is less than 120k.
4) The input of DPN is the entire image so that it could integrate more contextual information than patch-level segmentation models. Meantime, our method only needs one forward pass to generate the complete segmentation maps, thereby the inference speed of our method is faster than patch-level models.

\textit{Relationship with HRNet}~\citep{wang2020deep}.
Both DPN and HRNet learn a high-resolution representation, while there are some differences. 1) Proposed DPN only maintains one kind of resolution, i.e., the high-resolution stream (as can be seen in Fig.~\ref{fig:unet_vs_dpn}), and
multi-scale feature fusion is employed in DP-Blocks.
While HRNet maintains multi-streams with different resolutions and multi-scale fusion is employed among these streams. 2) The method for multi-scale feature fusion is different. Proposed DPN fuses low-resolution feature maps with high-resolution feature maps in a step-by-step manner. While in HRNet, the representations with different resolutions are concatenated directly.

\subsection{Detail-Preserving Block}
\begin{figure}
\centering
\includegraphics[width=0.40\textwidth]{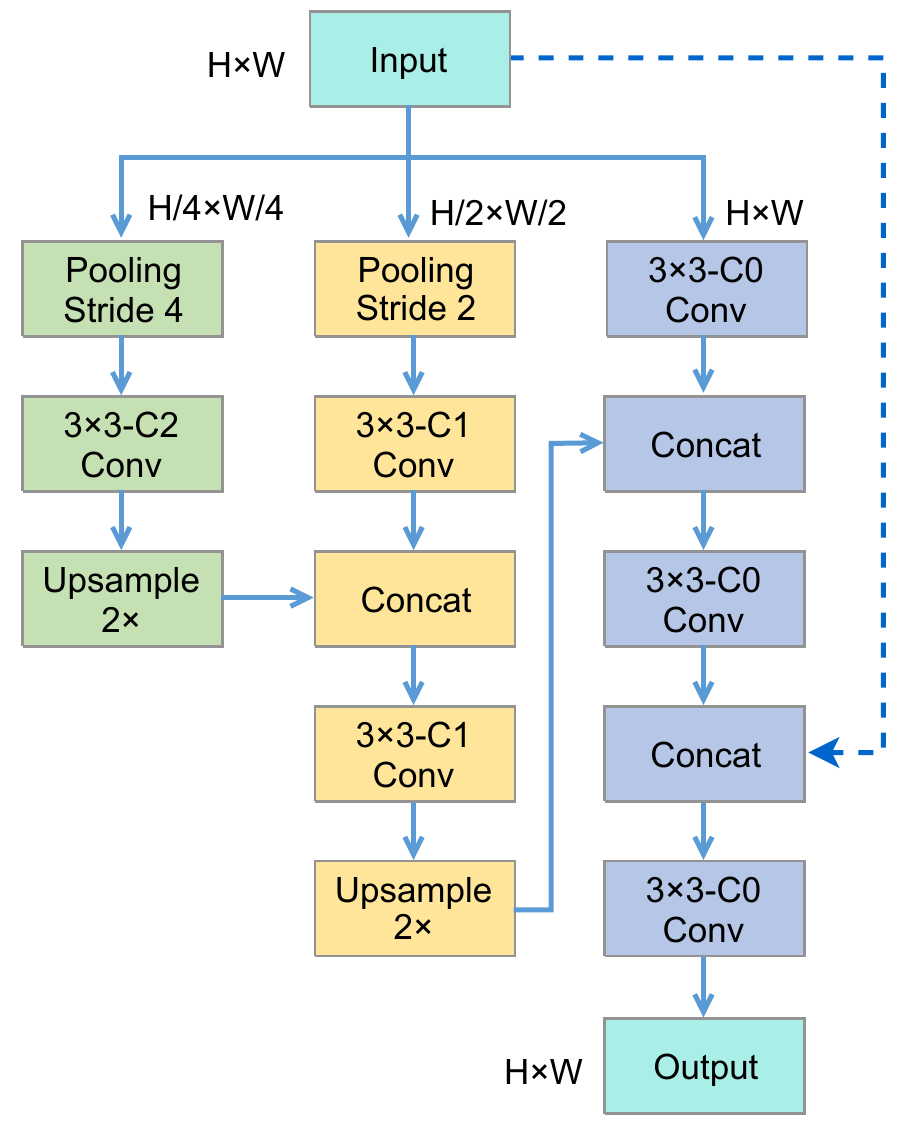}
\caption{Overview of the proposed detail-preserving block, where C0, C1, and C2 denote the number of convolutional filters for each branch.}
\label{fig:dpblock}
\end{figure}

DP-Block as the key component of DPN, could learn structural semantic information and preserve the spatial detailed information at the same time.
Overview of the DP-Block is visualized in Fig.~\ref{fig:dpblock}.
We can observe that the input feature of the DP-Block is fed into three branches, and each branch is processed in different scales.
The output feature of the DP-Block is obtained by fusing features of three scales. The computing procedure of the DP-Block is as follows.

For the first branch, a convolution operation with 3$\times$3 kernel was adopted to learn detailed information.
For the second branch, a pooling operation with stride 2 was adopted, then the resolution of feature maps was down-sampled by a factor of 2. A convolution operation with 3$\times$3 kernel was adopted further.
For the third branch, it was used to enlarge the field of view and learn structural information.
In this branch, a pooling operation with stride 4 was first adopted, as a result, the resolution of feature maps was down-sampled by a factor of 4 and the receptive field was increased by a factor of 4 either. A convolution operation with 3$\times$3 kernel was then adopted to extract features.
The extracted features of each branch were fused in a cascaded manner. Specifically, features learned by the third branch were first up-sampled 2$\times$, and then connected to the second branch, and the output of the second branch was further connected to the first branch. Here, we used a concatenation operation for feature fusion. The whole procedure is summarized in Alg~\ref{alg:algorithm1}.
We note that the resolution of the output feature of the DP-Block is the same as the input feature so that the DP-Block could not only preserve detailed information but also learns multi-scale features.

%insert algorithm description here
\begin{algorithm}
\caption{Description of the DP-Block.}
\label{alg:algorithm1}
	\KwIn{Feature map $X$}
	\KwOut{Feature map $Y$}
	\BlankLine
	
    $x1 = \delta(k1 * X)$;

    $x2 = \delta(k2 * maxpool(X,2))$;

    $x3 = \delta(k3 * maxpool(X,4))$;

    $x4 = \delta(k4 * concat(x2, deconv(x3,2)))$;

    $x5 = \delta(k5 * concat(x1, deconv(x4,2)))$;

    $Y = \delta(k6 * concat(x5, X))$;

     In the above formulas, $\delta$ refers to ReLU function, $*$ denotes convolution operation, $k_i$ denotes convolutional filters with kernel size 3$\times$3, $maxpool$ denotes max pooling, $deconv$ denotes deconvolution operation to upscale the feature map, and $concat$ denotes concatenation operation on the channel dimension.
\end{algorithm}

\textit{Number of parameters}. In our experiments, the number of convolutional filters C0, C1, and C2 for each branch of DP-Block was set to 16, 8, and 8, respectively. In DPN, the dimension of the output feature of the first convolution operation is H$\times$W$\times$32, then the number of parameters of the first DP-Block is 21,704.
For the second to the last DP-Block, the dimension of the input feature is H$\times$W$\times$16, then the number of parameters for each DP-Block is only 13,896. Hence, the total number of parameters of the DPN is less than 120k.
Experimental results show that the DPN could be effectively learned from scratch with limited training samples.

\textit{Relationship with Inception Module}. Different from the inception module~\citep{googlenet} that uses parallel convolution operations with different convolutional kernels to learn multi-scale features, our DP-Block adopts down-sampling first, so that the receptive field is further enlarged. The receptive field of each output neuron is four times that of the input neuron. As a result, the receptive field grows exponentially when stacking multiple DP-Blocks.
Furthermore, rather than parallel processing branches in the inception module, the features of different branches were fused in a cascaded manner in DP-Block to better learn multi-scale features.

\subsection{Loss Function}
Blood vessels account only for a small proportion of the entire image. Specifically, the proportion of vessels is 8.69\%/6.93\%/7.71\% on the DRIVE/CHASE\_DB1/HRF datasets, respectively. There exists a class imbalance problem in vessel segmentation. To solve this problem, we adopted class balanced cross-entropy loss~\citep{xie2017holistically}, which uses a weight factor to balance vessel pixels and non-vessel pixels.
The class-balanced cross-entropy loss is defined as follows.
\begin{equation}
L(p,y|\theta) = -\beta\sum\limits_{y_j = 1}\log{p_j} -(1-\beta)\sum\limits_{y_j=0 }\log{(1-p_j)}
\label{equ:bceloss}
\end{equation}
where $p$ is a probability map obtained by a sigmoid operation, and
$p_j$ denotes the probability that the $j^{th}$ pixel belongs to vessel.
In addition, $y$ denotes the ground truth, and $\theta$ denotes model parameters.
Rather than using a fixed value, the weight factor $\beta$ is calculated at each iteration based on the distribution of vessel pixels and non-vessel pixels.
The weight factor $\beta$ is defined as below.
\begin{equation}
\beta = \frac{N_-}{N_+ + N_-}
\label{equ:bceloss_beta}
\end{equation}
where $N_+$ denotes the number of vessel pixels, and $N_-$ denotes the number of non-vessel pixels. Since $N_- > N_+$ , the weight for vessel pixels is large than the weight for non-vessel pixels. So that the model would focus more on vessel pixels than non-vessel pixels.

Besides the segmentation loss after the last layer of DPN, we add three auxiliary losses to the intermediate layers of DPN to pass extra gradient signals to alleviate the gradient-vanish problem, just as that did in DSN~\citep{Lee2015DSN} and GoogLeNet~\citep{googlenet}.
As can be seen in Fig.~\ref{fig:net_overview}, the first auxiliary loss is after DP-Block2, the second auxiliary loss is after the DP-Block4, and the last one is after the DP-Block6. The segmentation loss is connected after the DP-Block8.
Taking the first auxiliary loss as an example, we first adopted a convolution operation with one 1$\times$1 filter to the output features of  DP-Block2, then a feature map with one channel was obtained. At last, this feature map was fed into the class balanced cross-entropy loss function.

Hence, the overall objective function of DPN is the sum of three auxiliary losses and one segmentation loss, and it can be formulated as follows.
\begin{equation}
L_{all}(x,y|\theta) = \sum_{i=1}^4{L(p^i(x),y|\theta)}+\frac{\lambda}{2}{||\theta||^2}
\label{equ:bceloss_beta}
\end{equation}
where $p^i$ denotes the probability map of the $i^{th}$ loss function, and $\lambda$ denotes the weight decay coefficient.

In conclusion, we aim to minimize the above objective function during training.
In the test phase, the output of the last segmentation loss is taken as the segmentation results of DPN, and the segmentation probability maps of auxiliary losses are ignored.
\section{Experiments}
\label{sec:exp}

\subsection{Materials}
Performances of our method were evaluated on three public datasets: DRIVE~\citep{drive}, CHASE\_DB1~\citep{chase}, and HRF~\citep{budai2013robust}.

The DRIVE (Digital Retinal Images for Vessel Extraction) dataset contains 40 color fundus images captured with a 45$^{\circ}$ FOV (Field of View).
Each image has the same resolution, which is 565$\times$584 (width$\times$height).
The dataset is partitioned into the training set and the test set officially, and each set contains 20 images.
For the test set, two groups of annotations are provided. We used the annotation of the first group as ground-truth to evaluate our model, just as other methods did. In addition, the FOV masks for calculating evaluation metrics are also provided.

The CHASE\_DB1 dataset contains 28 fundus images (999$\times$960) captured with a 30$^{\circ}$ FOV.
As the split of the training set and the test set is not provided. For a fair comparison with other methods, we did two sets of experiments.
We adopted a 20/8 partition for the first set of experiments, where the first 20 images were selected for training and the rest 8 images for testing. For another set of experiments, we adopted a 14/14 (training/test) partition.

The HRF (High Resolution Fundus) dataset contains 45 high-resolution fundus images, with resolution 2336$\times$3504. Among 45 images, 15 images are with diabetic retinopathy, 15 images are with glaucoma and the rest 15 images are healthy. As no partition of the training set and test set available, in our experiments, we used the first 15 images as the training set and the rest 30 images for evaluation. Besides, the FOV masks are provided in HRF.

\begin{figure}
\centering
\subfigure{
\includegraphics[width=0.2\textwidth]{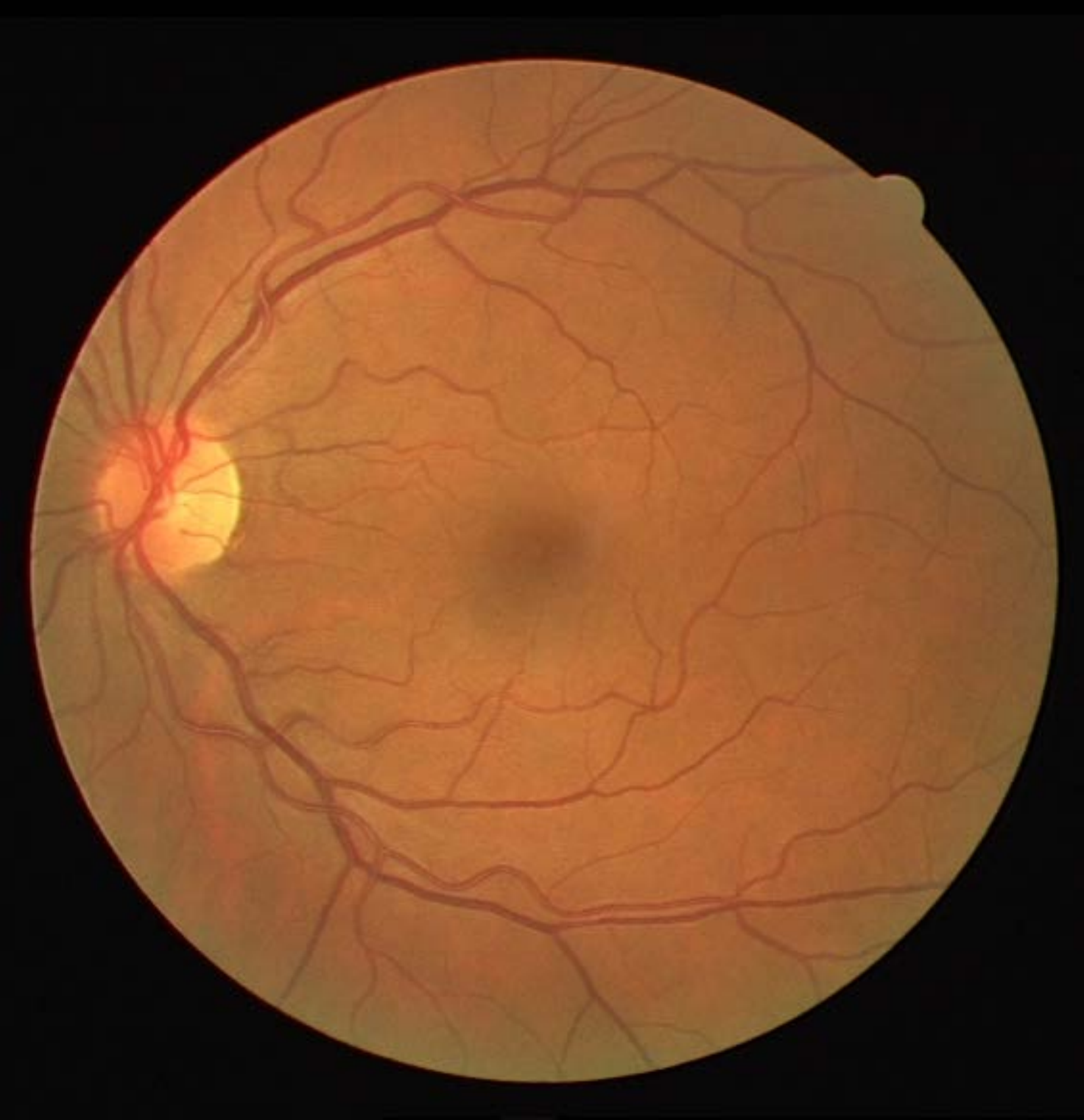}
}
\subfigure{
\includegraphics[width=0.2\textwidth]{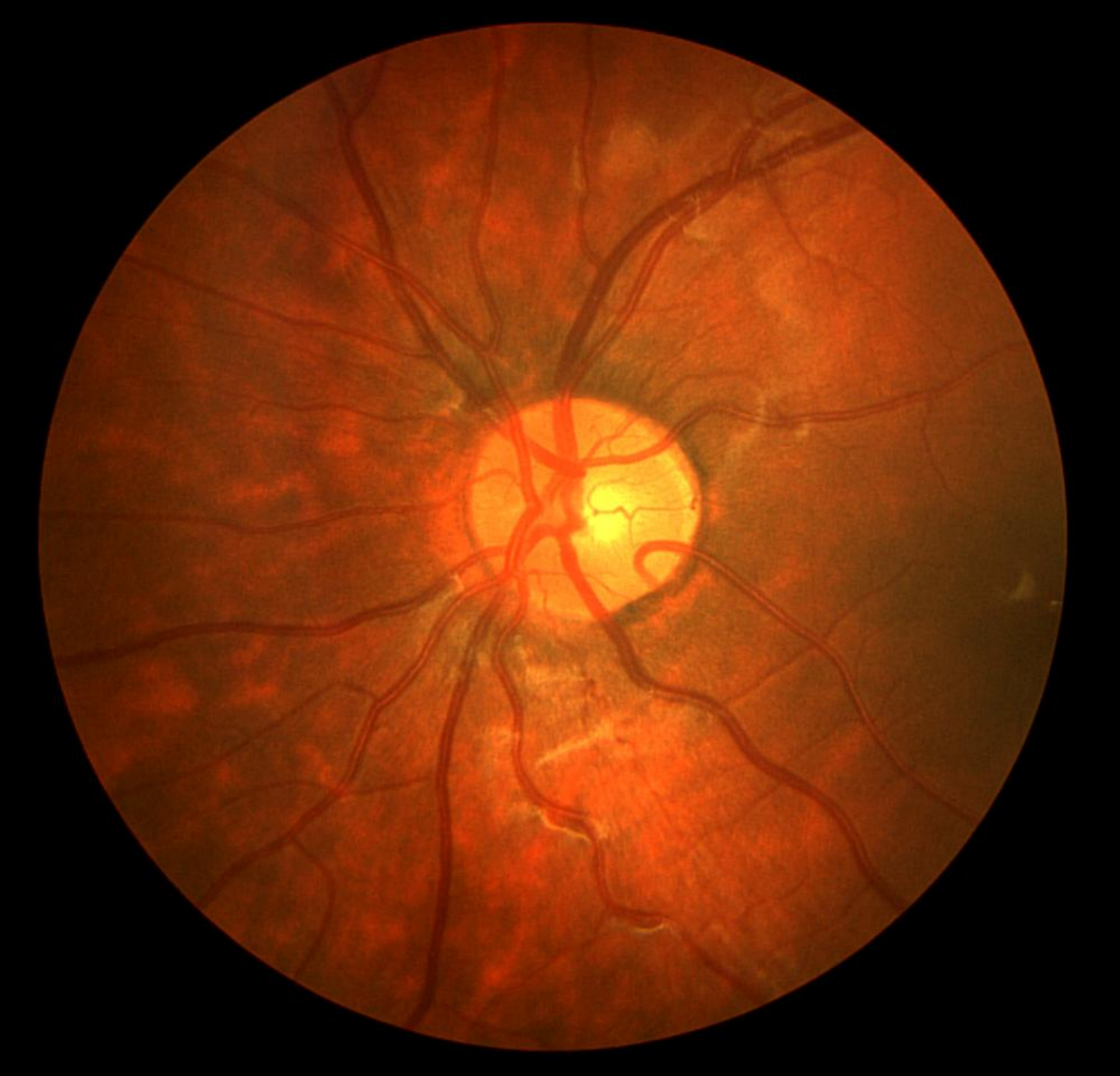}
}
\subfigure{
\includegraphics[width=0.28\textwidth]{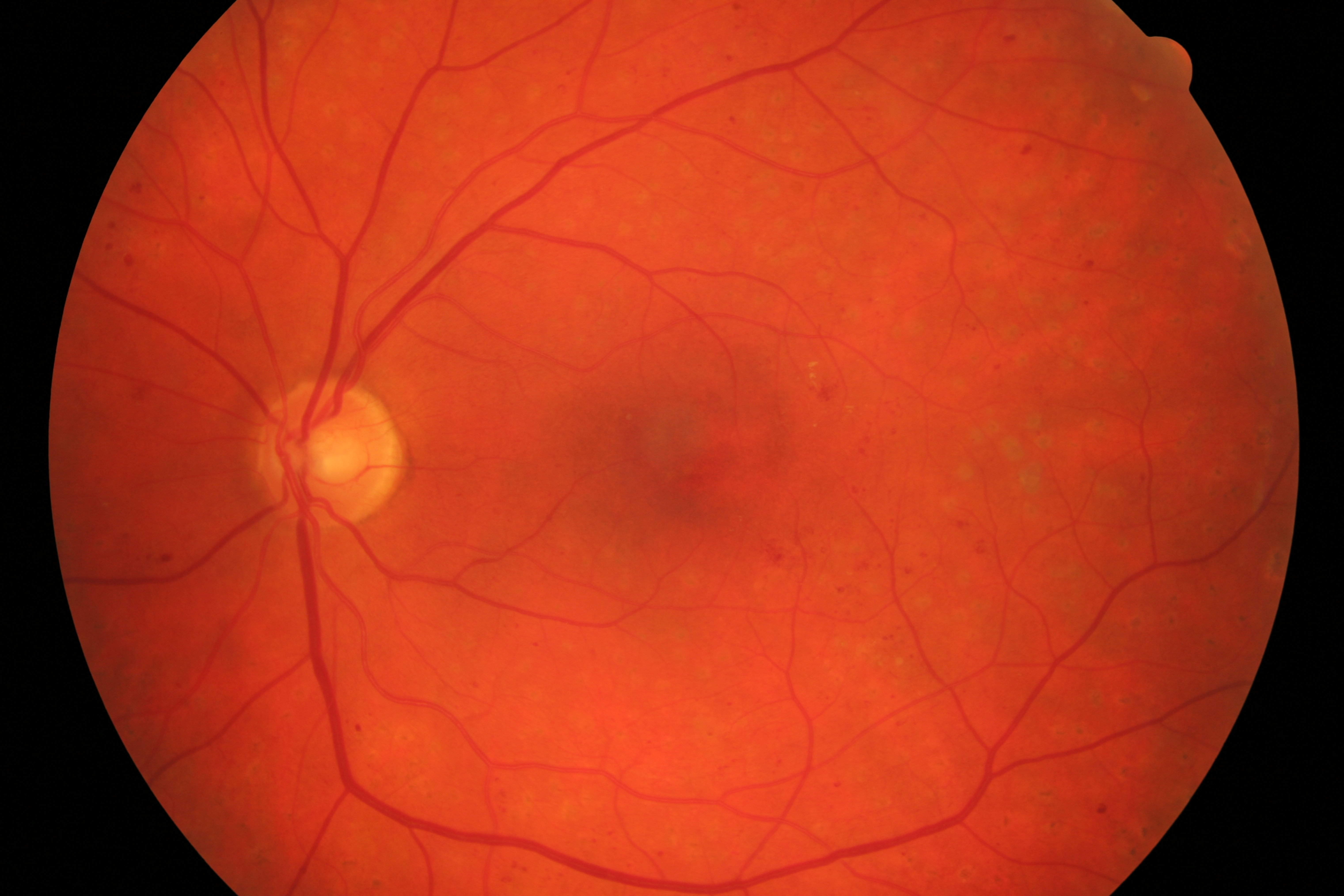}
}\\
\subfigure{
\includegraphics[width=0.2\textwidth]{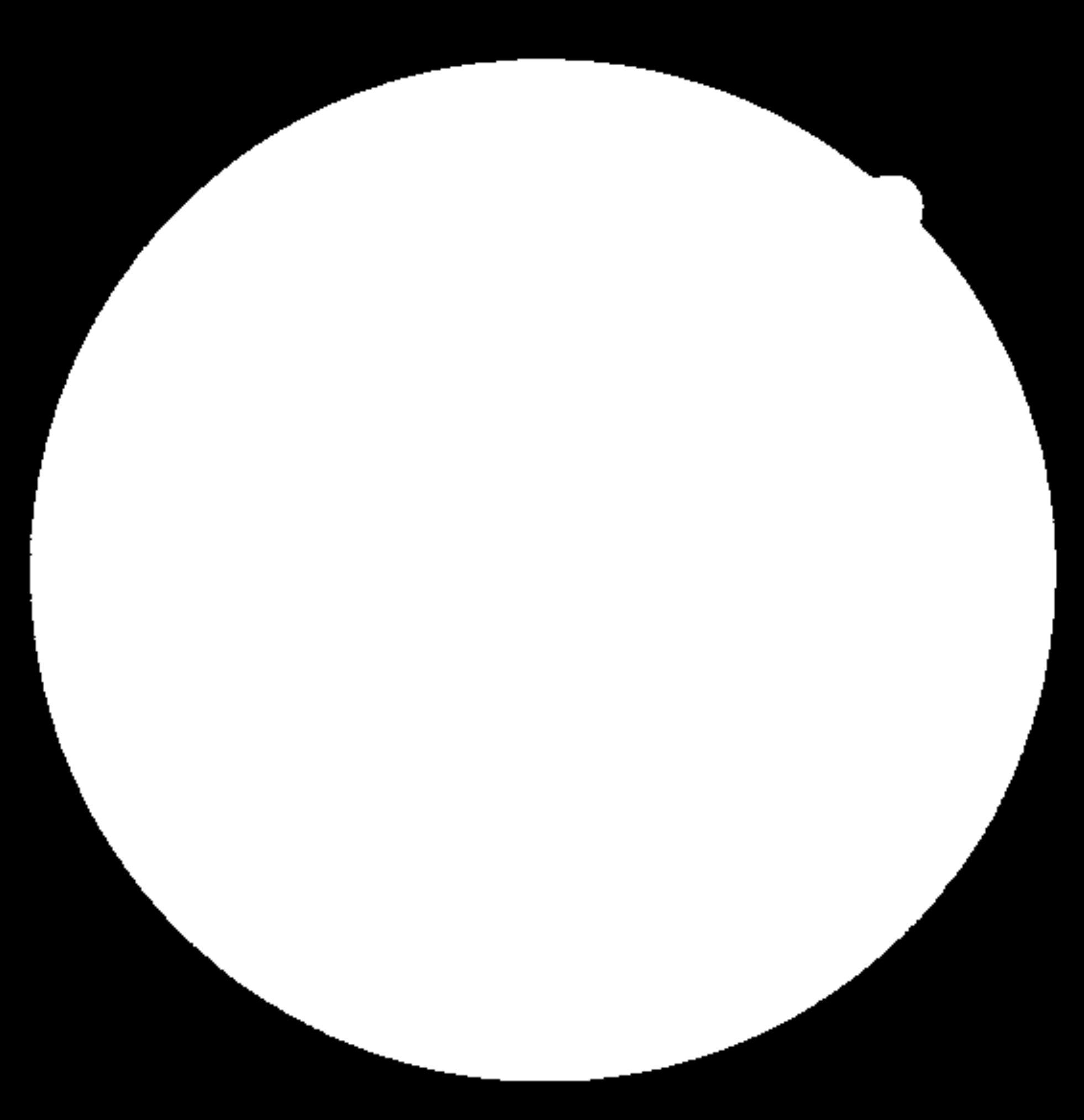}
}
\subfigure{
\includegraphics[width=0.2\textwidth]{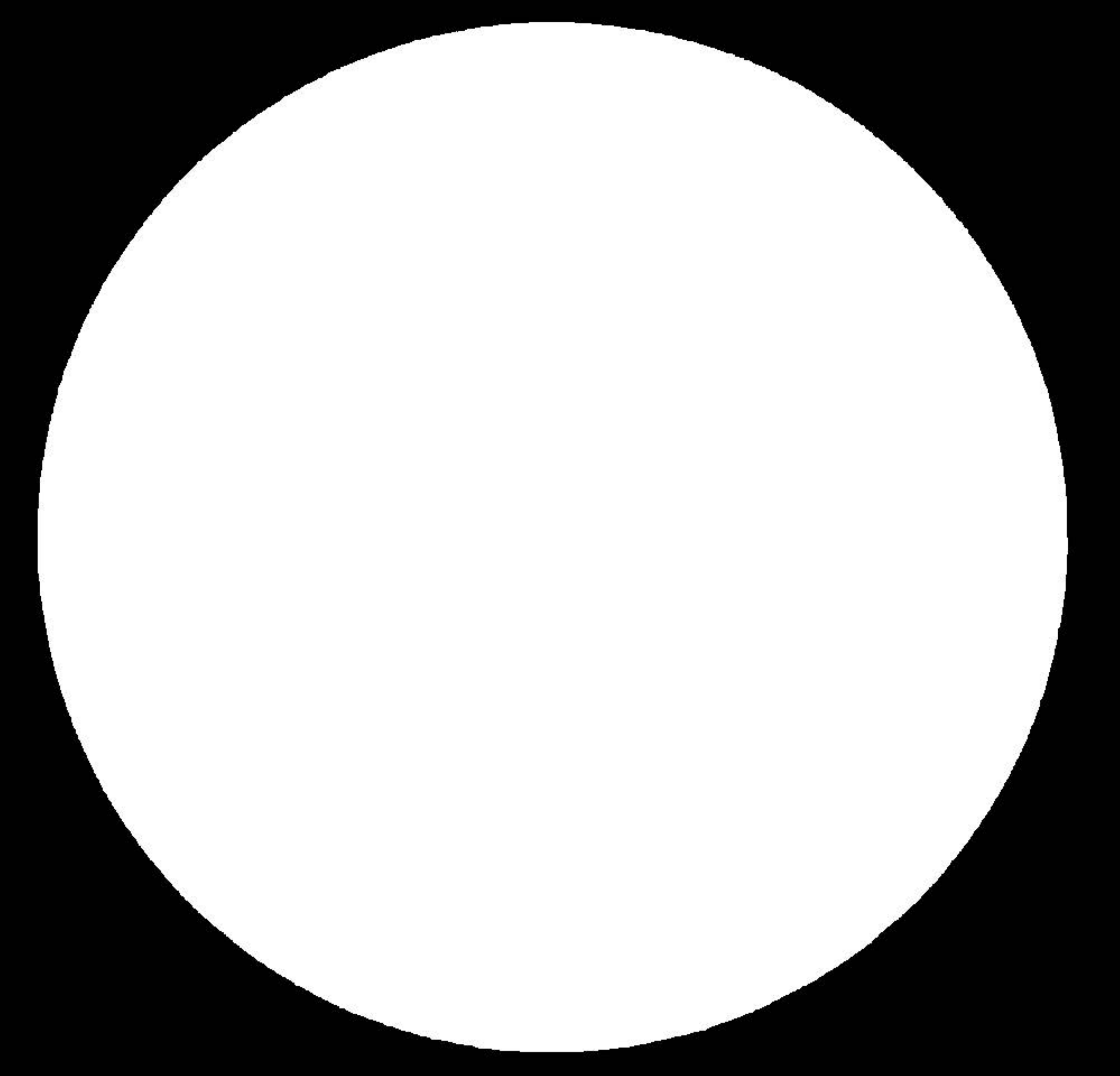}
}
\subfigure{
\includegraphics[width=0.28\textwidth]{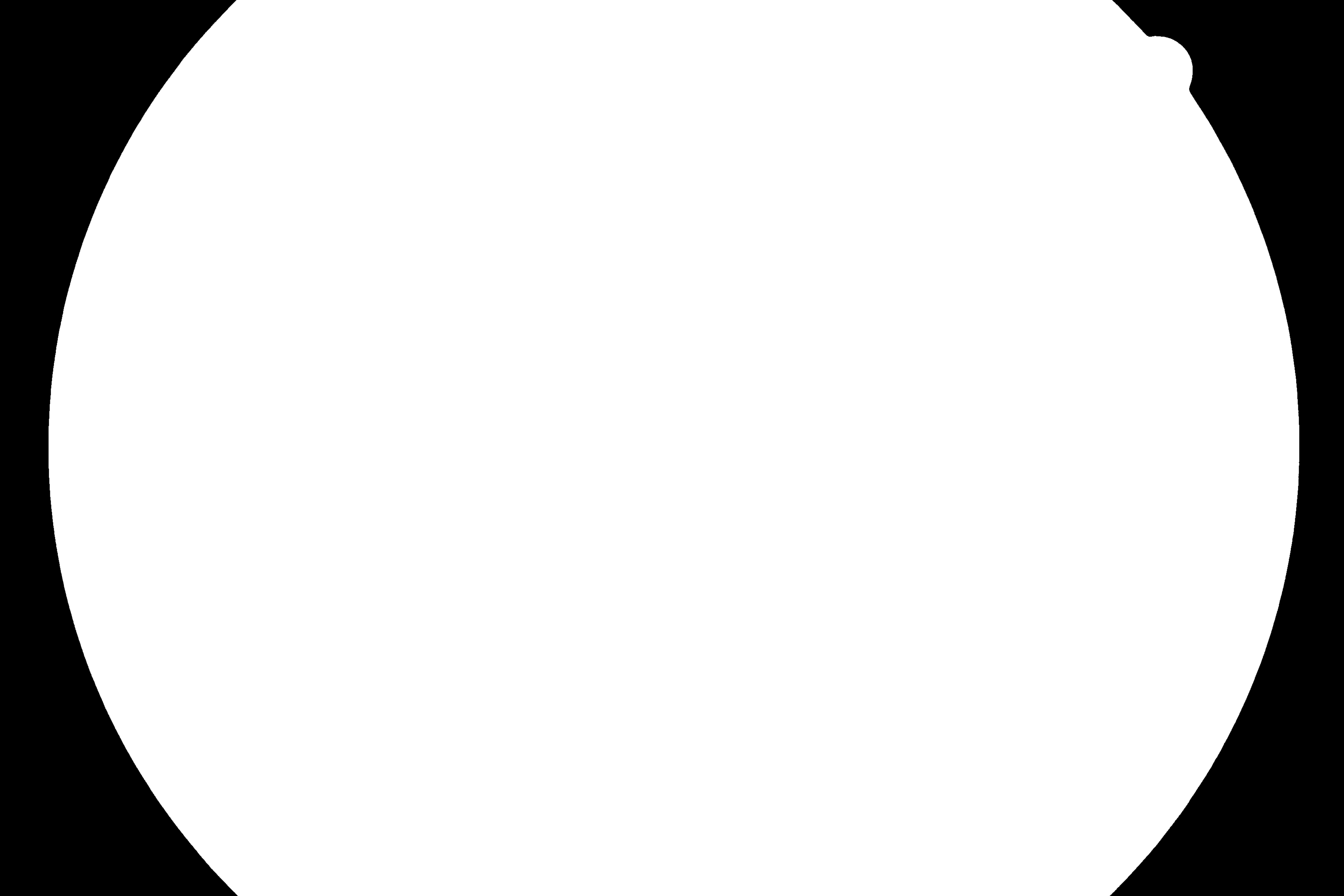}
}
\caption{Fundus images (the first row) and the corresponding FOV masks (the second row) from DRIVE, CHASE\_DB1 and HRF datasets, from left to right.}
\label{fig:fov_mask}
\end{figure}

As the FOV masks are not present in the CHASE\_DB1 dataset, we created the masks manually. The FOV masks on these three datasets are presented in Fig.~\ref{fig:fov_mask}.

\subsection{Image Preparation}
We use raw color fundus images to train our model, image enhancement methods like CLAHE are not adopted. Therefore, a time-consuming pre-processing procedure could be avoided when inference and the segmentation speed of our model could be further improved.

\begin{figure}
\centering
\subfigure[0$^{\circ}$]{
\includegraphics[width=0.12\textwidth]{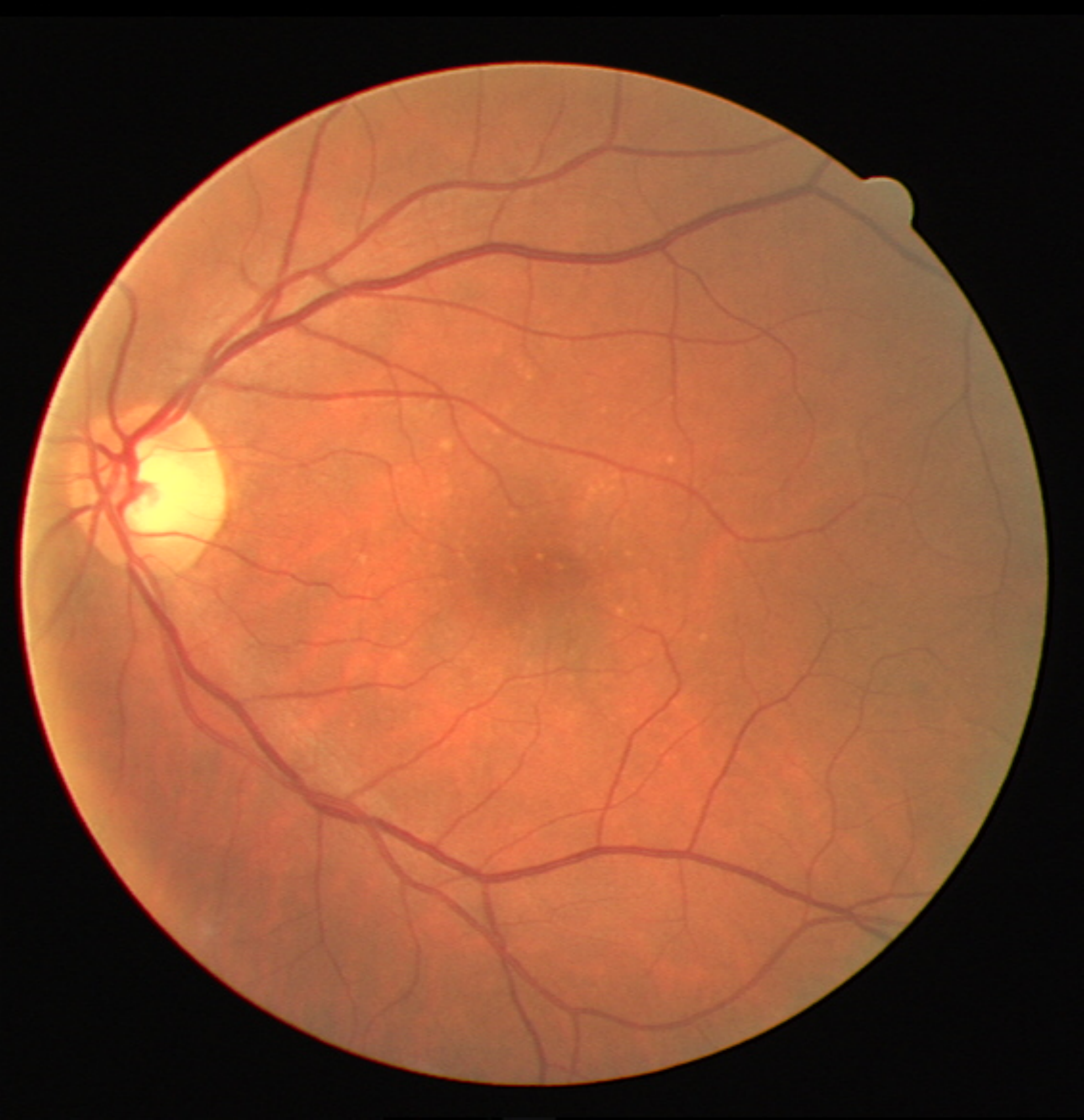}
}
\subfigure[22$^{\circ}$]{
\includegraphics[width=0.12\textwidth]{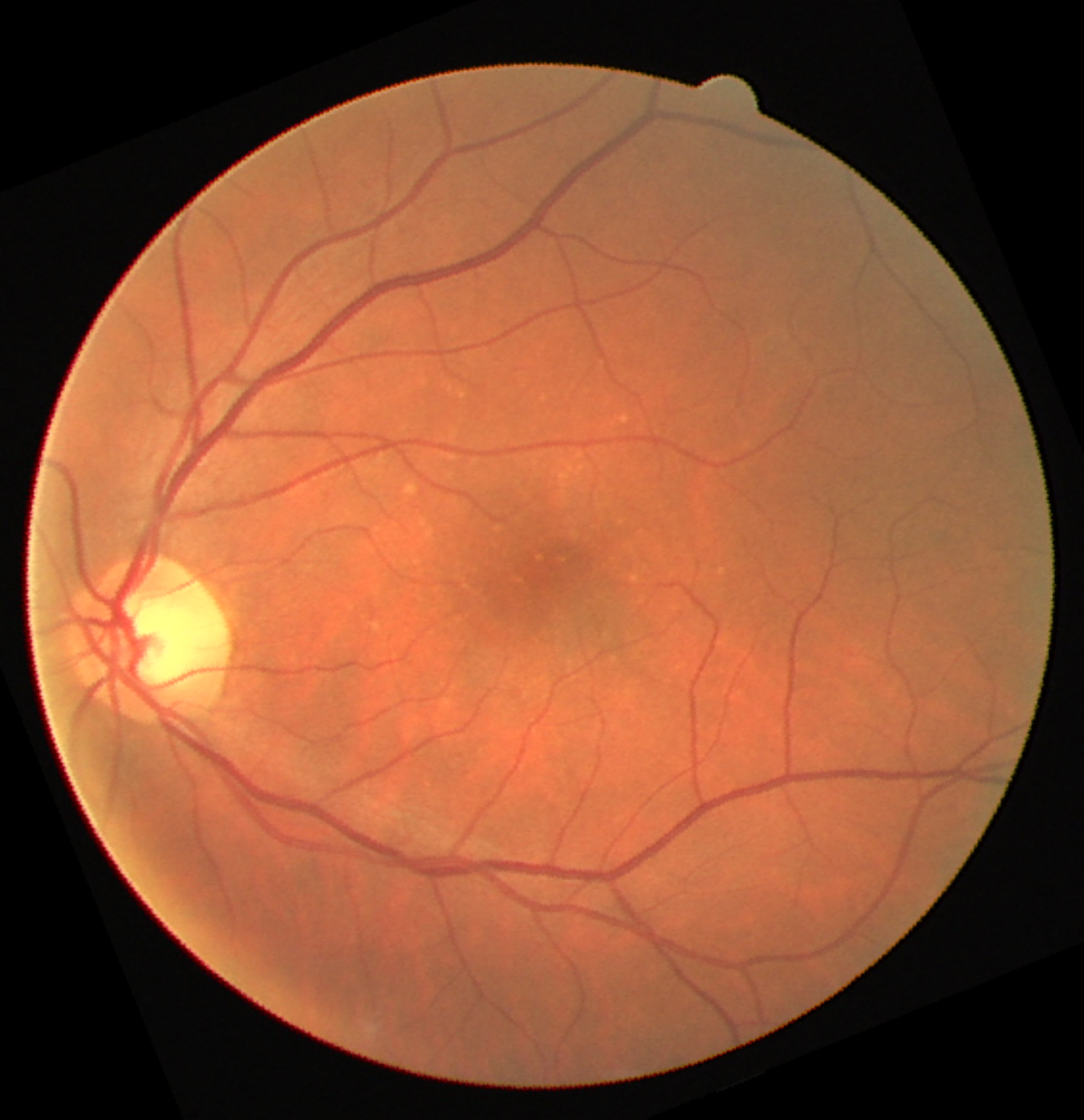}
}
\subfigure[45$^{\circ}$]{
\includegraphics[width=0.12\textwidth]{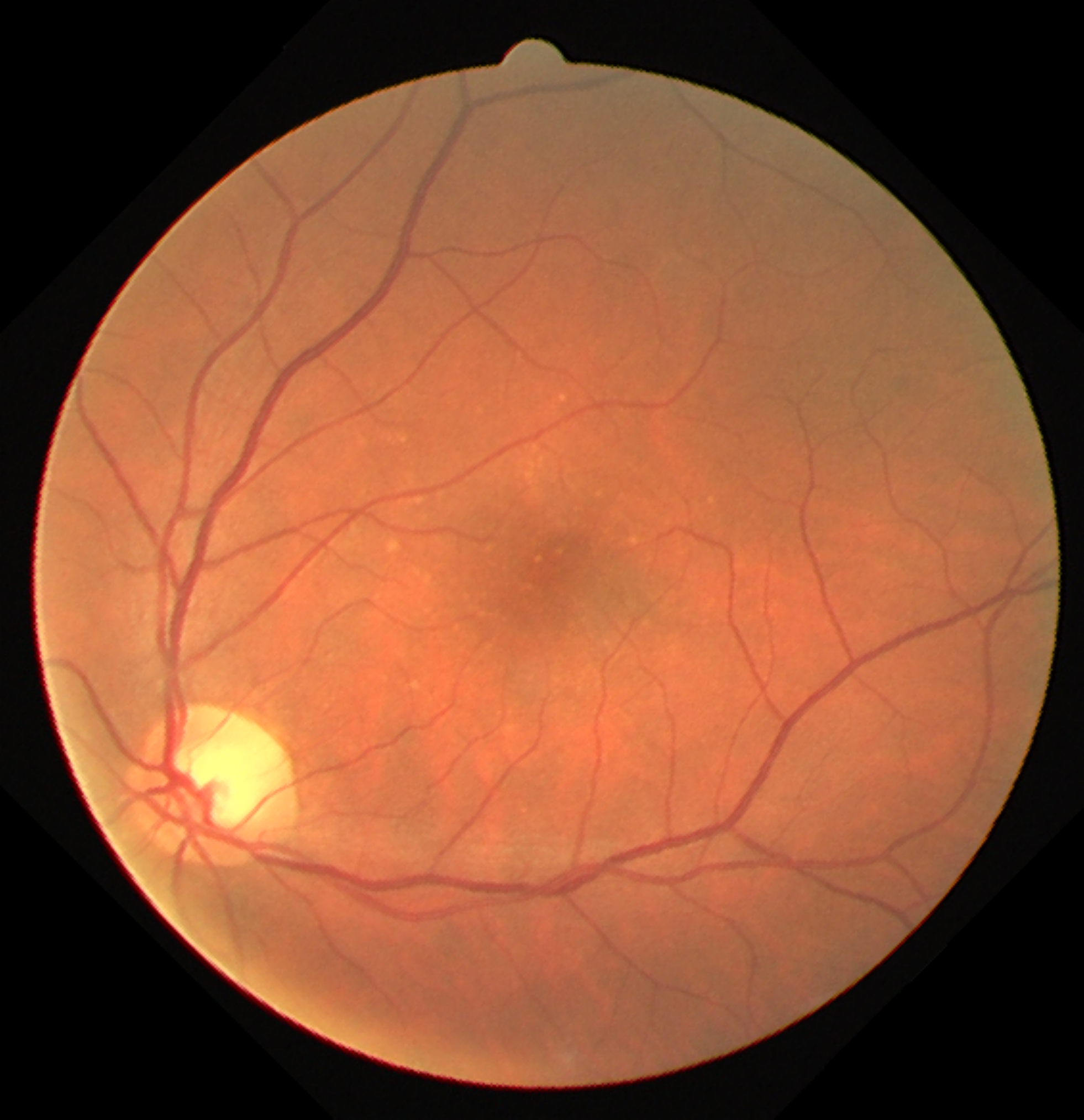}
}
\subfigure[90$^{\circ}$]{
\includegraphics[width=0.12\textwidth]{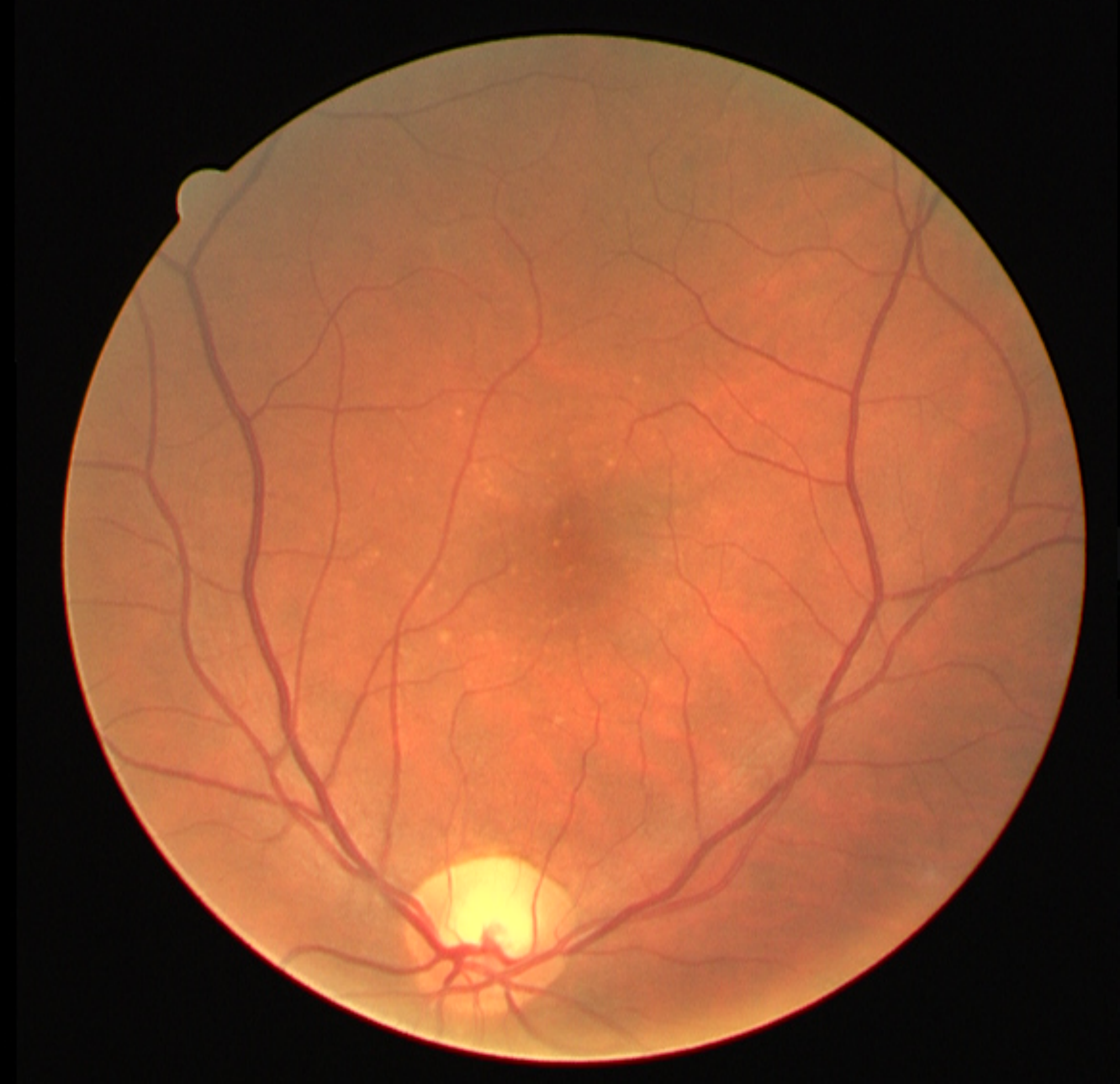}
}
\subfigure[135$^{\circ}$]{
\includegraphics[width=0.12\textwidth]{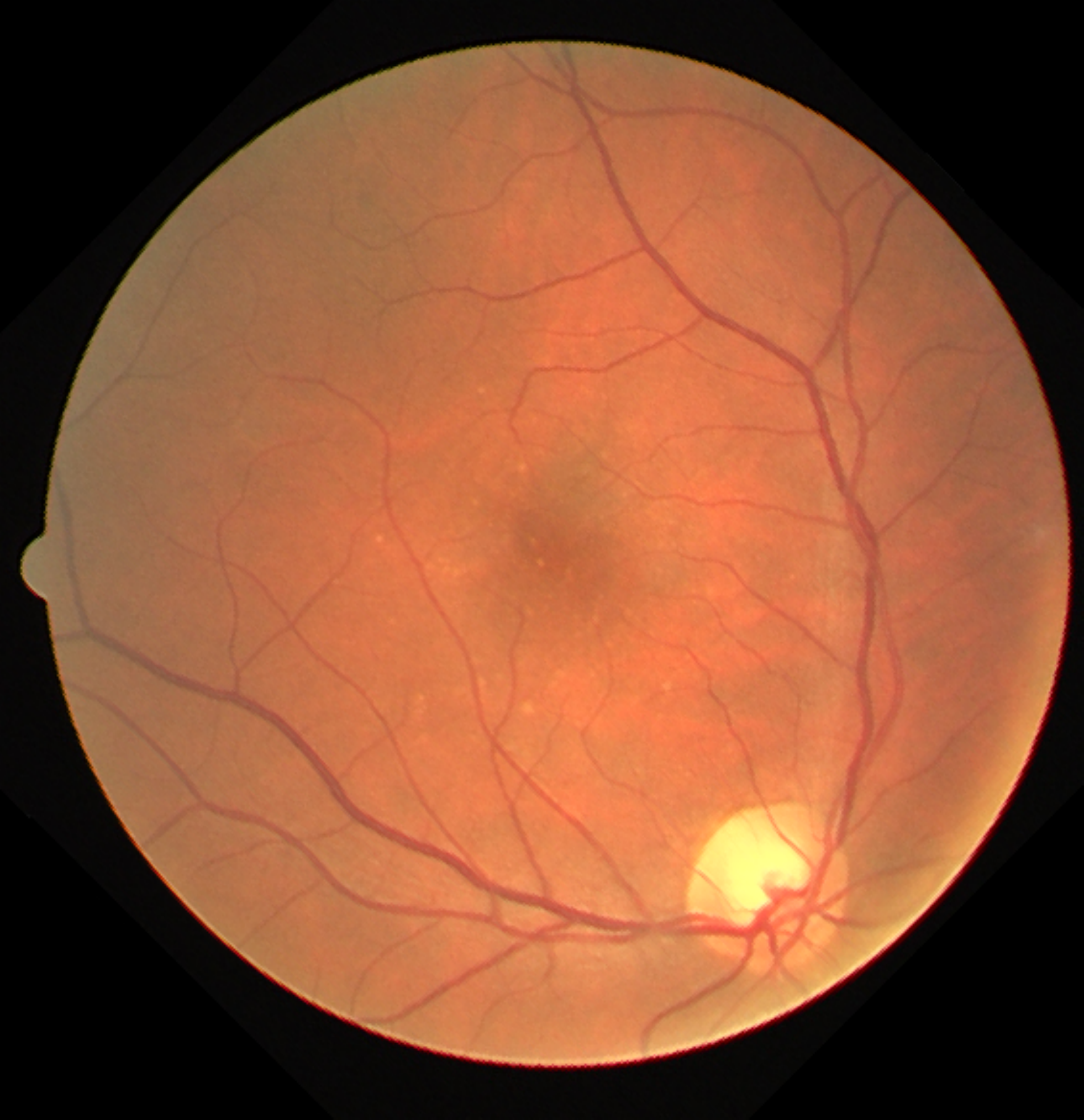}
}
\subfigure[180$^{\circ}$]{
\includegraphics[width=0.12\textwidth]{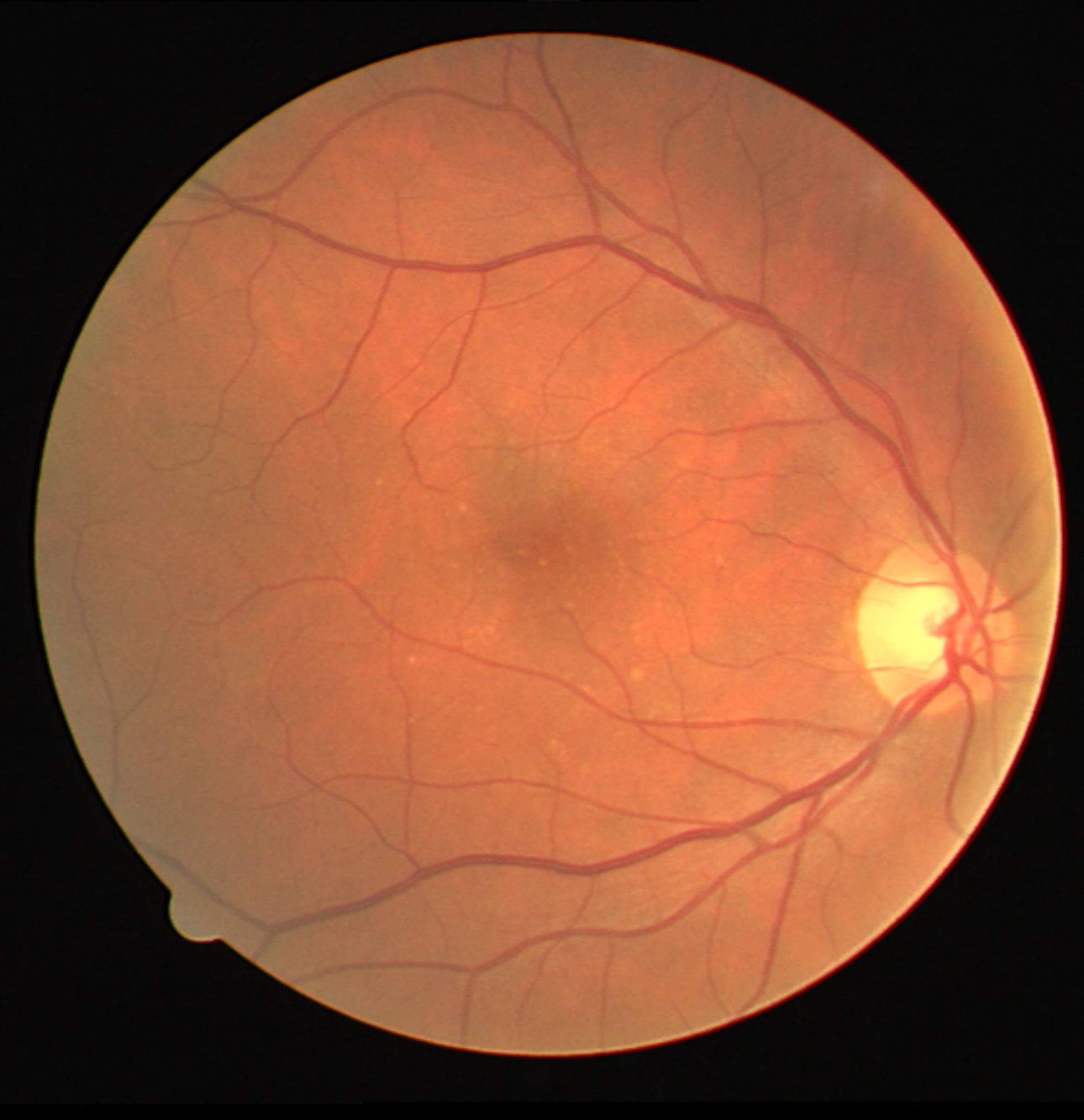}
}
\caption{Example of training images after rotating different angles.}
\label{fig:rot}
\end{figure}

Considering that there are limited annotated training samples on the DRIVE, CHASE\_DB1, and HRF, and there are no pre-training weights available.
To avoid over-fitting and promote the segmentation accuracy,
several transformations have been adopted to augment the training set, including flipping (horizontal and vertical) and rotation (22$^{\circ}$, 45$^{\circ}$, 90$^{\circ}$, 135$^{\circ}$, 180$^{\circ}$, 225$^{\circ}$, 270$^{\circ}$, 315$^{\circ}$). Rotated images can be seen in Fig.~\ref{fig:rot}.
As a result, the training images were augmented by a factor of 10 offline,
and there are 220/220/165 training images in total on the DRIVE/CHASE\_DB1/HRF datasets, respectively.
Moreover, the training image was randomly mirrored during training for each iteration.

On the DRIVE and CHASE\_DB1 datasets, it is possible to feed an entire image into GPU memory.
However, it is hard to feed an image sampled from HRF into GPU memory, since its resolution is as high as 2336$\times$3504. To deal with this problem, we scale these images from HRF to a low-resolution with size 600$\times$900 to make sure they can be loaded into GPU memory.

\subsection{Training Details}
Our model was implemented based on an open-source deep learning framework \textit{Caffe}~\citep{caffe}, and it ran on a workstation equipped with one NVIDIA RTX 2080ti GPU.

We initialized weights of our model using xavier~\citep{glorot2010understanding}. The learning rate was initialized to 1e-3. And we trained our model for 100k/100k/70k iterations with ADAM~\citep{adam} (batch size 1) using weight decay 0.0005 on the DRIVE/CHASE\_DB1/HRF datasets, respectively.

To reduce computational complexity, each training image was cropped into 512$\times$512/632$\times$632/588$\times$588 patches randomly during training on the DRIVE/CHASE\_DB1/HRF datasets, respectively.
And, the crop operation was performed via the data layer of~\textit{Caffe}. When testing, the entire fundus image is fed into the network without cropping, so that our model could generate a segmentation map with only one forward pass.

\subsection{Evaluation Metrics}
For a fair comparison with other methods~\cite{jin2019dunet,wang2019dual,driu,GUO2019BTS}, we adopt five pixel-wise evaluation metrics, including Sensitivity (Se), Specificity (Sp), Accuracy (Acc), the Area Under the receiving operator characteristics Curve (AUC), and F1\textrm{-}score (F1) to evaluate the segmentation maps. Besides these pixel-wise evaluation metrics, we also use Structural Similarity Index Measure (SSIM)~\cite{wang2004image} and Peak Signal to Noise Ratio (PSNR) to evaluate the segmentation maps.
They are defined as follows.

\begin{align}
Se = \frac{TP}{TP+FN}
\end{align}

\begin{align}
Sp=\frac{TN}{TN+FP}
\end{align}

\begin{align}
Acc=\frac{TP+TN}{TP+FN+TN+FP}
\end{align}

\begin{align}
F1=\frac{2\times Pr\times Se}{Pr+Se}
\end{align}

where $Pr = \frac{TP}{TP+FP}$, and true positive (TP) denotes the number of vessel pixels classified correctly and true negative (TN) denotes the number of non-vessel pixels classified correctly. Similarly, false positive (FP) denotes the number of non-vessel pixels misclassified as vessels and false negative (FN) denotes the number of vessel pixels misclassified as non-vessels.
To calculate Se, Sp and Acc, we select the threshold that corresponds to the optimal operating point of the receiving operator
characteristics (ROC) curve to generate the binary segmentation maps from a probability map.
Also, we note that TP, FN, FP and TN are counted pixel-by-pixel, and only the pixels inside the FOV mask are calculated, not the whole fundus image.
The ROC curve is obtained by  multiple Se versus (1-Sp) via varying threshold. AUC evaluates the segmentation probability maps not the binary maps, which is more comprehensive.
The AUC ranges from 0 to 1, and the AUC of a perfect segmentation model is 1.

At last, we also report the segmentation speed of our model using fps (frames per second). The segmentation time $t$ for each image is counted starting from reading the raw test image from the hard disk to writing the segmentation map into the hard disk. Then, $fps = 1.0 / t$.

\subsection{Results and Analysis}

\subsubsection{Compare with Existing Methods}
We compared our method with several state-of-the-art deep vessel segmentation methods on three public datasets in terms of segmentation performance, segmentation speed, and the number of parameters.
Comparison results were summarized in Table~\ref{table:drive_vel_result}, Table~\ref{table:chase_vel_result} and Table~\ref{table:hrf_vel_result}.

\begin{table}
\setlength{\tabcolsep}{1.5pt}
\tiny{
\begin{center}
\caption{Comparison results on the DRIVE dataset (For each metric, the best results are shown in bold.)}
\begin{threeparttable}
\begin{tabular}{ccccccccccc}
\hline
Method &One Forward Pass? &Se & Sp &Acc &AUC &F1 &SSIM &PSNR &fps &Params(M)\\
\hline
Liskowski et al.~\citep{Liskowski2016Segmenting} &No
&0.7811 &0.9807 &0.9535 &0.9790 &N.A &N.A &N.A  &N.A &N.A\\
FCN~\citep{OLIVEIRA2018229}&No      &\textbf{0.8039}  &0.9804   &\tgr{0.9576}  &\textbf{0.9821}  &N.A         &N.A &N.A &\tbl{0.5}     &\tgr{0.2}\\
U-Net~\citep{jin2019dunet} &No      &0.7849        &0.9802   &0.9554        &0.9761        &0.8175      &N.A &N.A &0.32    &3.4\\
DUNet~\citep{jin2019dunet} &No      &0.7963        &0.9800   &0.9566        &0.9802        &0.8237      &N.A &N.A &0.07    &0.9\\
DEU-Net~\citep{wang2019dual}&No    &0.7940        &\tbl{0.9816}&0.9567        &0.9772        &\tgr{0.8270}&N.A &N.A &0.15    &N.A \\
MS-NFN~\citep{wu2018}      &No      &0.7844        &\tgr{0.9819}&0.9567     &0.9807        &N.A         &N.A &N.A &0.1 &\tbl{0.4} \\
Patch BTS-DSN~\citep{GUO2019BTS}&No &0.7891       &0.9804       &0.9561     &0.9806        &\tbl{0.8249} &0.5159 &13.5640  &N.A  &7.8\\
Three-stage FCN~\citep{threestage}&No&0.7631      &\textbf{0.9820} &0.9538     &0.9750        &N.A &N.A  &N.A  &N.A &20.4\\
Vessel-Net~\citep{wu2019vessel}&No  &\tgr{0.8038} &0.9802    &\textbf{0.9578}  &\textbf{0.9821}  &N.A &N.A  &N.A &N.A  &1.7\\
DRIU~\citep{driu}         &Yes     &0.7855        &0.9799    &0.9552        &0.9793        &0.8220  &0.5923 &13.6364 &N.A     &7.8\\
Image BTS-DSN~\citep{GUO2019BTS} &Yes&0.7800      &0.9806 &0.9551     &0.9796        &0.8208      &\textbf{0.5941} &\textbf{14.2633} &N.A   &7.8 \\
\hline
Our Method               &Yes   &0.7934 &0.9810  &0.9571
&0.9816 &\textbf{0.8289} &0.5500 &13.7672 &\textbf{11.8}  &\textbf{0.1}\\
\hline
\end{tabular}
\label{table:drive_vel_result}
\begin{tablenotes}
\item[1] N.A : Not Available
\end{tablenotes}
\end{threeparttable}
\end{center}
}
\end{table}

\begin{table*}
\setlength{\tabcolsep}{1.5pt}
\tiny{
\begin{center}
\caption{Comparison results on the CHASE\_DB1 dataset (For each metric, the best results are shown in bold.)}
\begin{threeparttable}
\begin{tabular}{ccccccccccc}
\hline
Method&One Forward Pass? &Se & Sp &Acc &AUC &F1 &SSIM &PSNR &fps &Split of dataset\\
\hline
MS-NFN~\citep{wu2018}      &No        &0.7538       &\textbf{0.9847} &0.9637       &0.9825       &N.A &N.A &N.A    &$<$0.1  &20/8 (train/test)\\
Three-stage FCN~\citep{threestage}&No &0.7641       &0.9806       &0.9607       &0.9776       &N.A &N.A &N.A  &N.A  &20/8 (train/test)\\
Vessel-Net~\citep{wu2019vessel}   &No &\textbf{0.8132} &0.9814       &\textbf{0.9661} &\textbf{0.9860} &N.A &N.A &N.A  &N.A  &20/8 (train/test)\\
Xu et al.~\citep{xu2020retinal} &No &N.A &N.A &0.9650 &0.9856 &N.A &N.A &N.A &N.A &20/8 (train/test)\\
DEU-Net~\citep{wang2019dual}&No       &\tgr{0.8074} &\tbl{0.9821} &\textbf{0.9661} &0.9812       &\tgr{0.8037} &N.A &N.A &0.08 &20/8 (train/test)\\
BTS-DSN~\citep{GUO2019BTS} &Yes       &\tbl{0.7888} &0.9801       &0.9627   &\tbl{0.9840} &\tbl{0.7983} &0.6052 &14.2717 &N.A &20/8 (train/test)\\
\hline
Our Method    &Yes       &0.7839 &0.9842 &0.9660 &\textbf{0.9860} &\textbf{0.8124}
&\textbf{0.6602} &\textbf{14.4583} &\textbf{5.6} &20/8 (train/test)\\
\hline
\hline
U-Net~\citep{jin2019dunet} &No         &0.8355 &0.9698 &0.9578 &0.9784   &0.7792 &N.A &N.A &0.10 &14/14 (train/test)\\
DUNet~\citep{jin2019dunet} &No         &0.8155 &0.9752 &0.9610 &0.9804   &0.7883 &N.A &N.A &0.02 &14/14 (train/test)\\
\hline
Our Method                &Yes        &0.7645 &\textbf{0.9846} &\textbf{0.9650} &\textbf{0.9840} &\textbf{0.8021} &0.6985 &14.7485 &\textbf{5.6}  &14/14 (train/test) \\
\hline
\end{tabular}
\label{table:chase_vel_result}
\begin{tablenotes}
\item[1] N.A : Not Available
\end{tablenotes}
\end{threeparttable}
\end{center}
}
\end{table*}

\begin{table}[!bt]
\begin{center}
\caption{Comparison results on the HRF dataset (For each metric, the best results are shown in bold.)}
\begin{threeparttable}
\begin{tabular}{ccccccccc}
\hline
Method &Se & Sp &Acc &AUC &F1 &SSIM &PSNR\\
\hline
Orlando et al.~\citep{orlando2016discriminatively}         &0.7874 &0.9584 &N.A     &N.A     &0.7158&N.A&N.A\\
Zhao et al.~\citep{zhao2017automatic}         &0.7490 &0.9420 &0.9410 &\textbf{0.9710} &N.A    &N.A&N.A\\
Yan et al.~\citep{yan2018joint}         &0.7881 &0.9592 &0.9437 &N.A  &N.A &N.A  &N.A    \\
\hline
Our Method &\textbf{0.7926} &\textbf{0.9764} &\textbf{0.9591} &0.9697 &\textbf{0.7835} &0.3062 &13.0284\\
\hline
\end{tabular}
\label{table:hrf_vel_result}
\begin{tablenotes}
\item[1] N.A : Not Available
\end{tablenotes}
\end{threeparttable}
\end{center}
\end{table}

\textit{Segmentation performance}.
On the DRIVE dataset, as we can see from Table~\ref{table:drive_vel_result},
our method achieves the highest F1-score compared with other methods, and the AUC is higher than 9 models. To be specific,
compared with DRIU~\citep{driu} and BTS-DSN~\citep{GUO2019BTS} which need only one forward pass to generate the segmentation map during testing, our method achieves much higher Se, Acc, AUC, and F1. Specifically, the Acc and AUC of our method are about 0.2\% higher than DRIU and BTS-DSN.
Besides, compared with the other eight methods that need multiple forward passes to generate the segmentation map during testing for only one fundus image, the Se, Acc, Auc, and F1 of our method is higher than six of the eight methods.

On the CHASE\_DB1 dataset, we compare our method with eight existing methods.
Our method achieves the highest AUC and F1 compared with other state-of-the-art methods, as can be seen in Table~\ref{table:chase_vel_result}. Besides, compared with U-Net and its variant models, namely DUNet and DEU-Net, our method shows superior performance in terms of Sp, AUC, and F1-score.

Besides adopting pixel-wise evaluation metrics, we also report the SSIM and PSNR of our method.
We can observe that the SSIM and PSNR of our method outperform BTS-DSN on the CHASE\_DB1 datasets by a significant margin. To be specific, the SSIM of DPN is over 5.5\% higher than that of BTS-DSN.
On the DRIVE dataset, the PSNR of our method is better than that of Patch BTS-DSN and DRIU, but lower than Image BTS-DSN. Although Image BTS-DSN achieves better SSIM and PSNR, its acc and AUC are lower than our method. We conclude that our method still shows competitive performance in terms of SSIM and PSNR.

At last, on the HRF dataset, our method also shows its superior performance compared with other methods. This part of the experiments real that our method achieves comparable or even superior performance in segmentation performance.

\textit{Segmentation speed}.
On the DRIVE dataset, the fps of our method is over 10, while the fps of all state-of-the-art methods are lower than 1.0.
For instance, the segmentation speed of our method is over 20$\times$, 100$\times$  faster than FCN~\citep{OLIVEIRA2018229} and MS-NFN~\citep{wu2018}, respectively.

On the CHASE\_DB1 dataset, most existing state-of-the methods require multiple forward passes and a recomposed operation to generate a segmentation map for one fundus image, thus they show slow segmentation speed.
Compared with DUNet~\citep{jin2019dunet} and DEU-Net~\citep{wang2019dual}, they need over 10s to segment a fundus image with resolution 999$\times$960. However, our method runs in an end-to-end way, and it could segment an image within 0.2s, which is over 280$\times$ and 70$\times$ faster than DUNet and DEU-Net.

We conclude that our method has obvious advantages in segmentation speed, which could better meet the real-time requirement in a clinical scene.

\textit{Number of parameters}.
The number of parameters of our model is only 120k, which is far less than all state-of-the-art models. Therefore, our method is more suitable to be deployed to edge devices due to its lightweight characteristic.

\subsubsection{Visualization}
\begin{figure*}
\centering
\subfigure{
\includegraphics[width=0.2\textwidth]{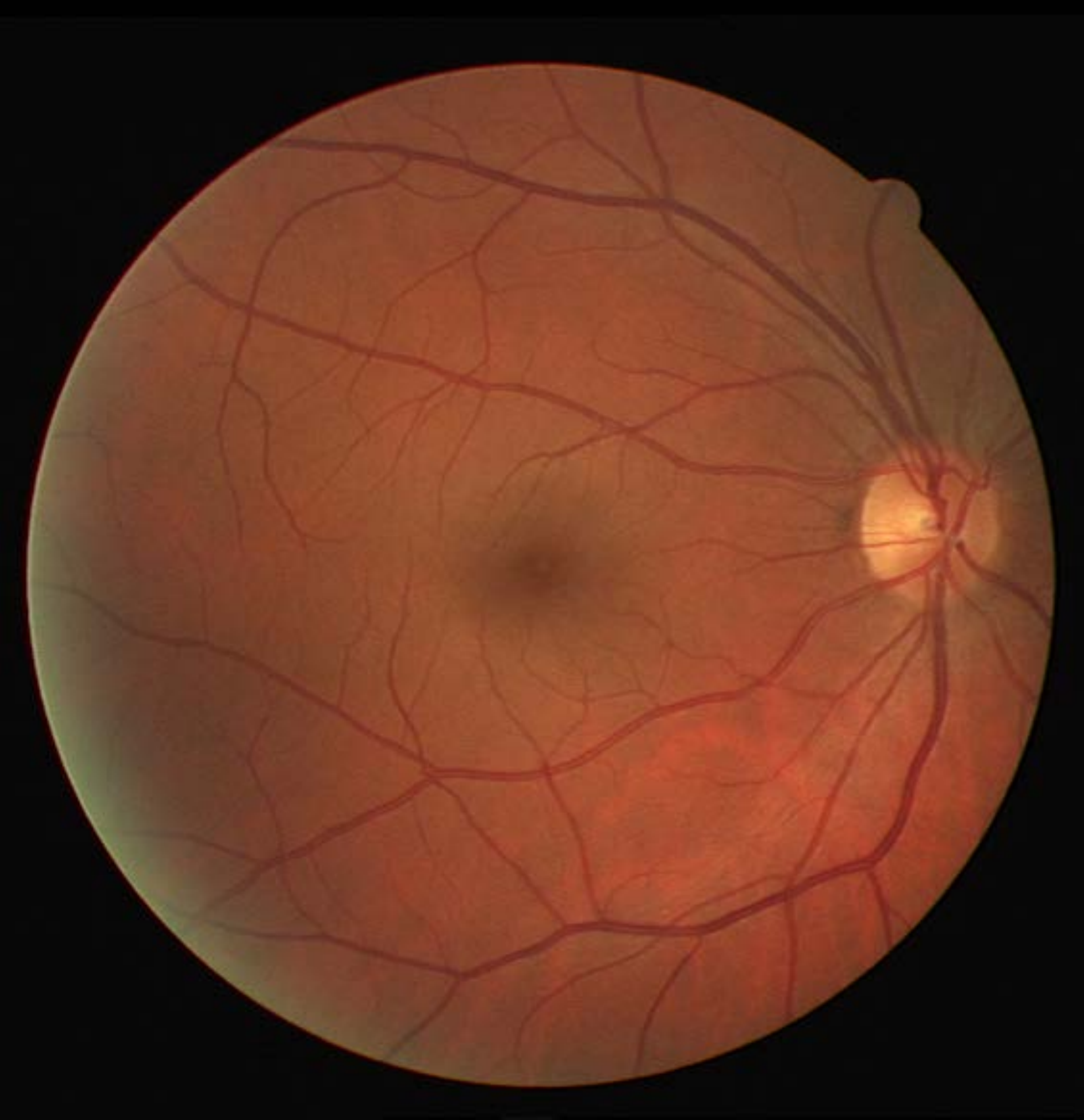}
}
\subfigure{
\includegraphics[width=0.2\textwidth]{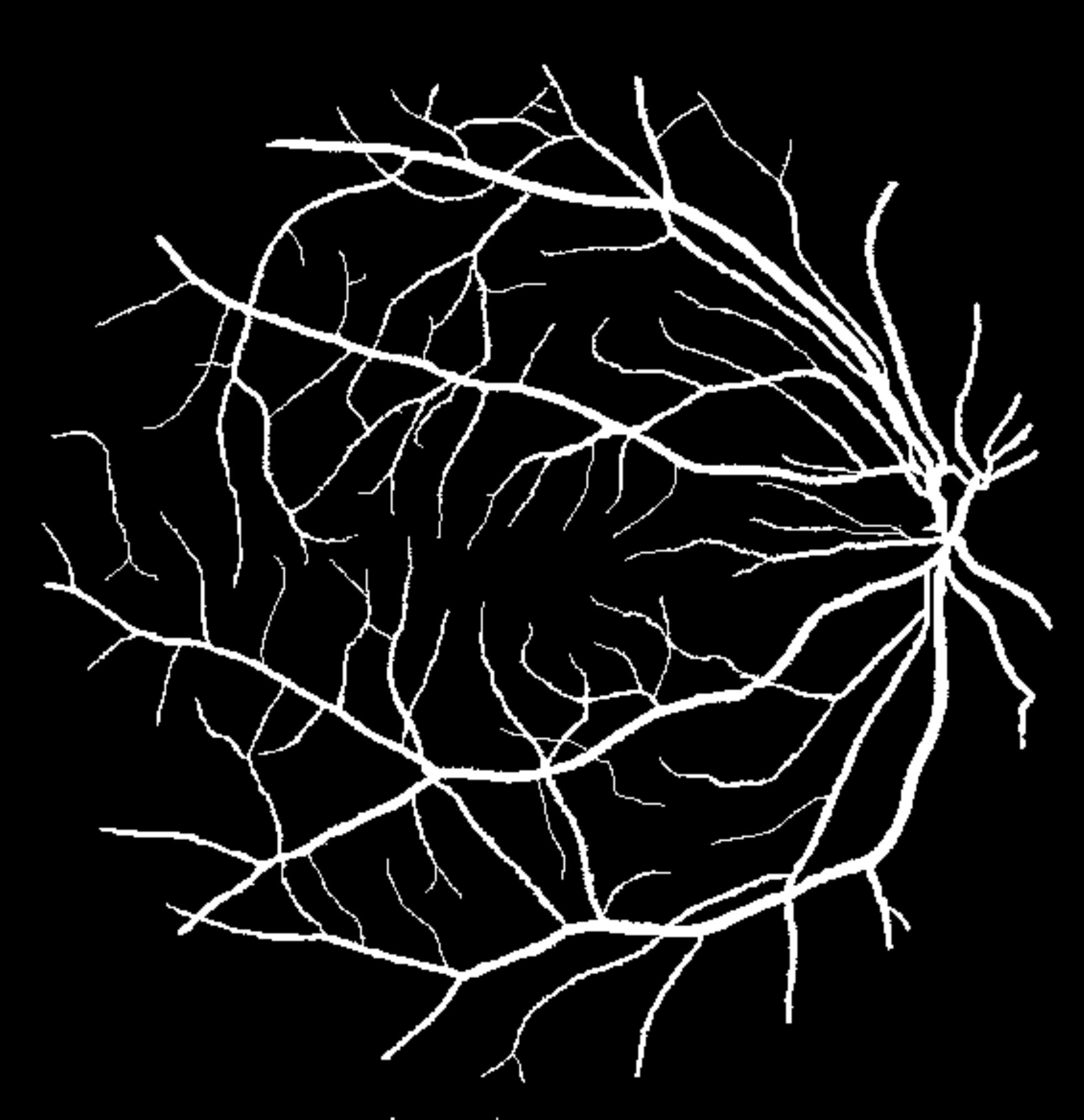}
}
\subfigure{
\includegraphics[width=0.2\textwidth]{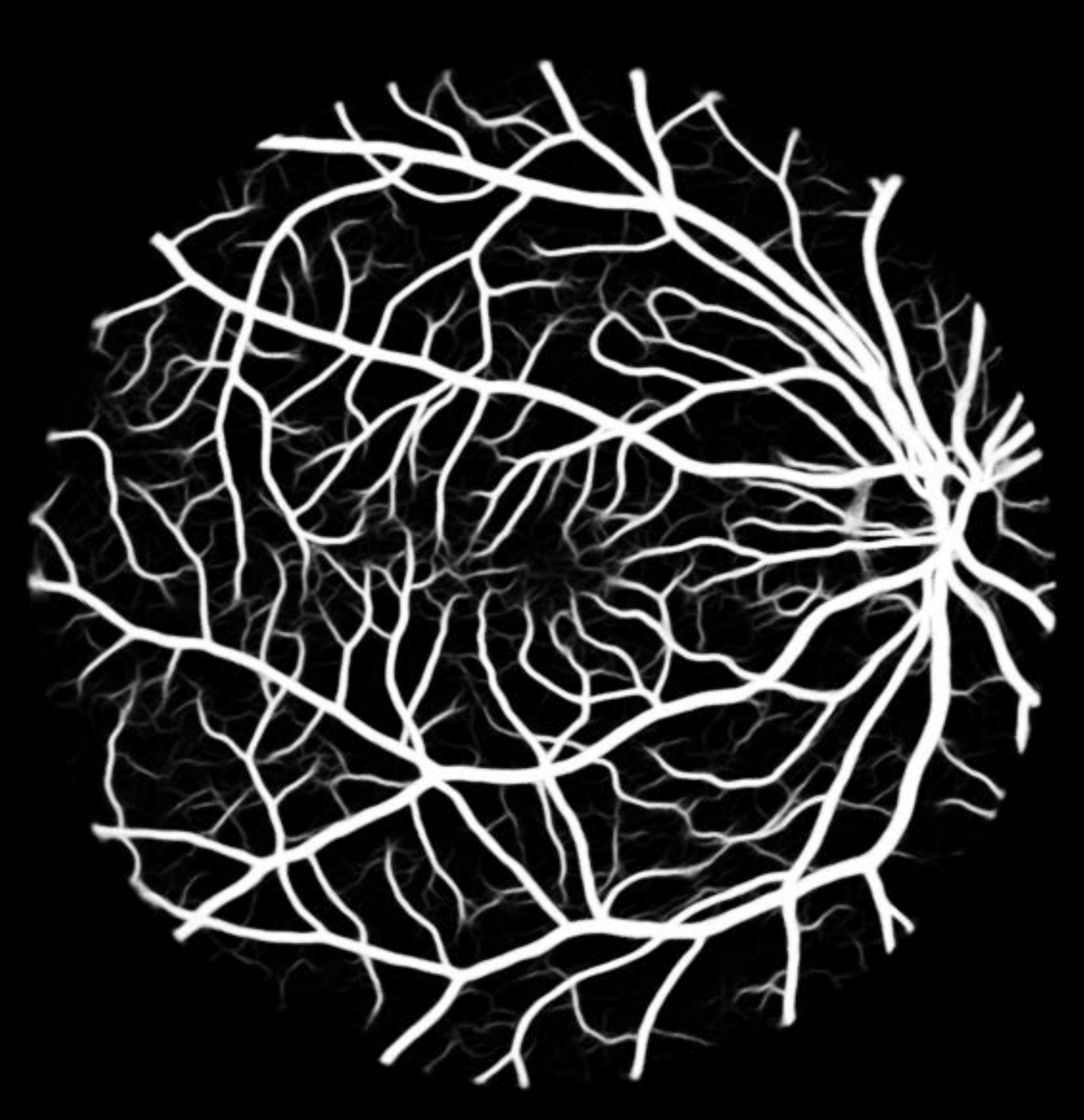}
}
\subfigure{
\includegraphics[width=0.2\textwidth]{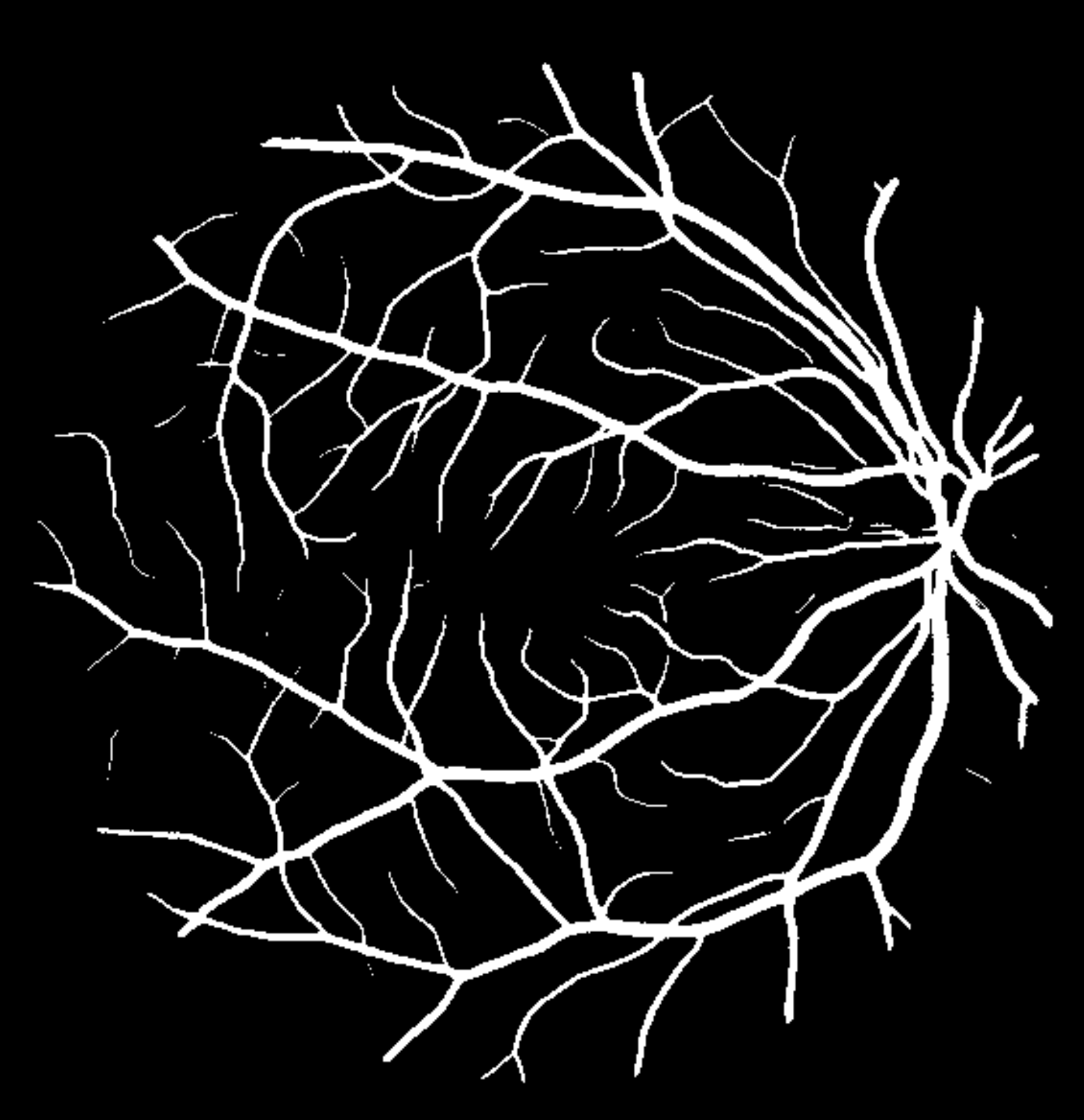}
}
\\
\subfigure{
\includegraphics[width=0.2\textwidth]{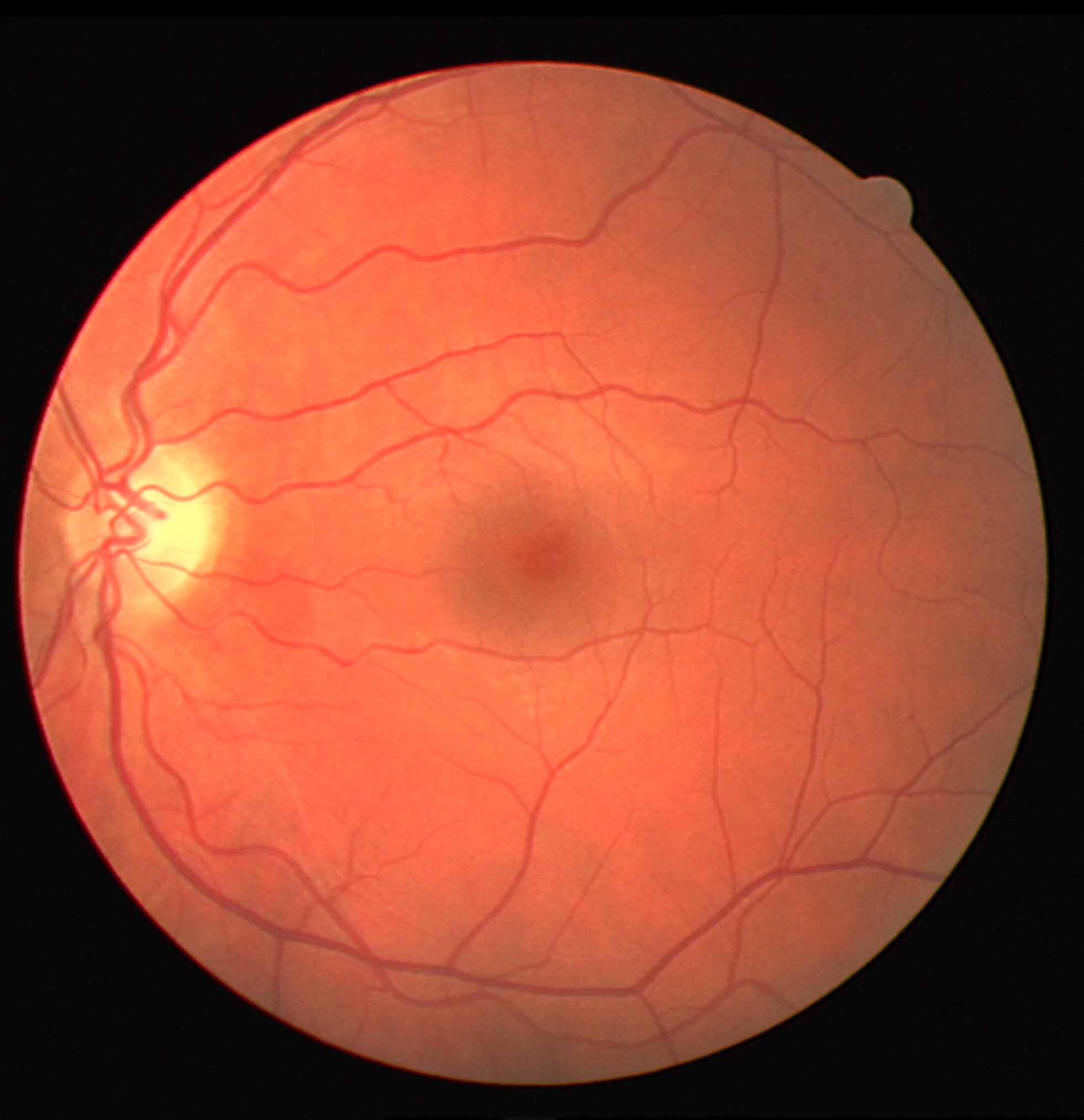}
}
\subfigure{
\includegraphics[width=0.2\textwidth]{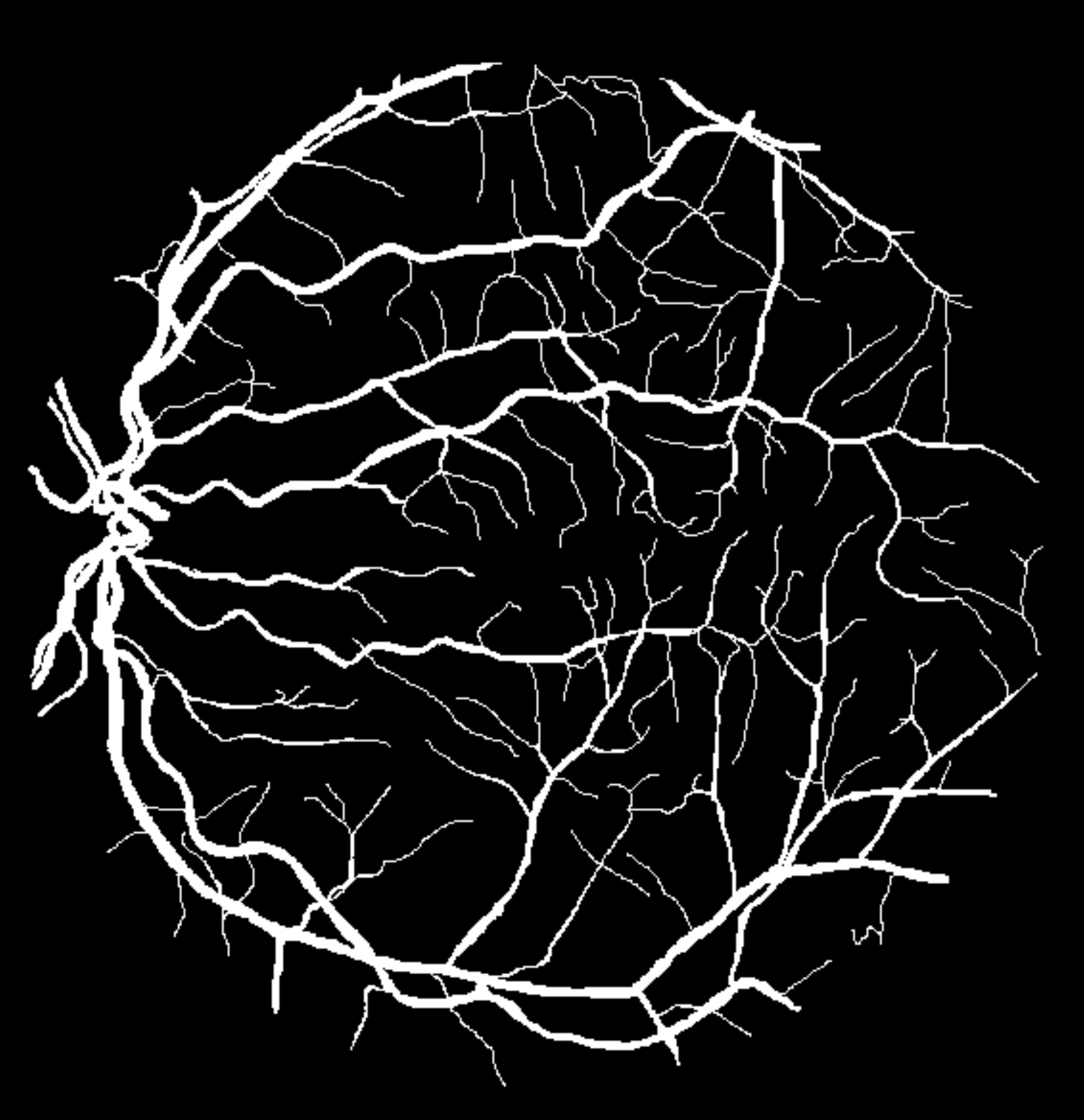}
}
\subfigure{
\includegraphics[width=0.2\textwidth]{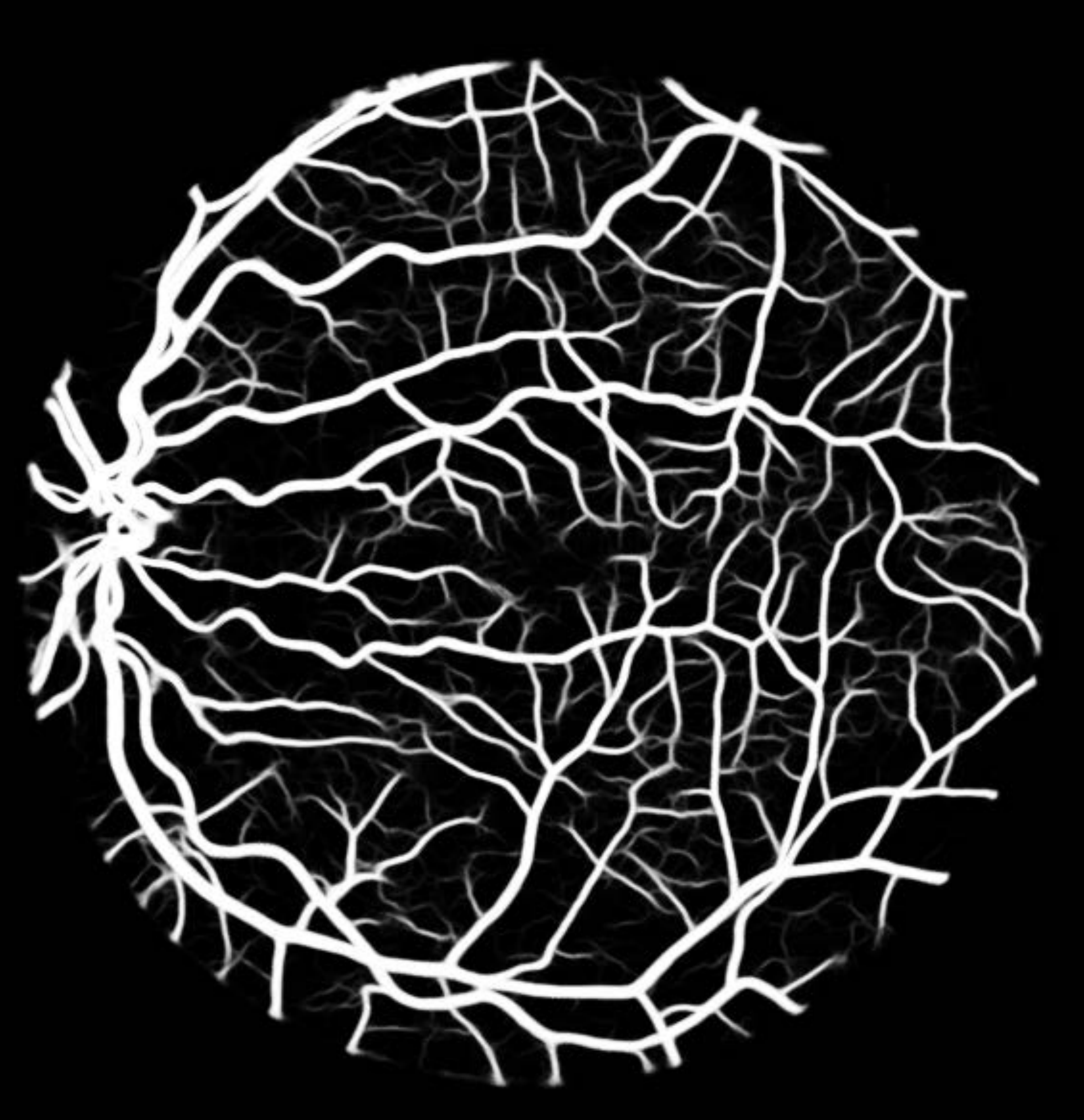}
}
\subfigure{
\includegraphics[width=0.2\textwidth]{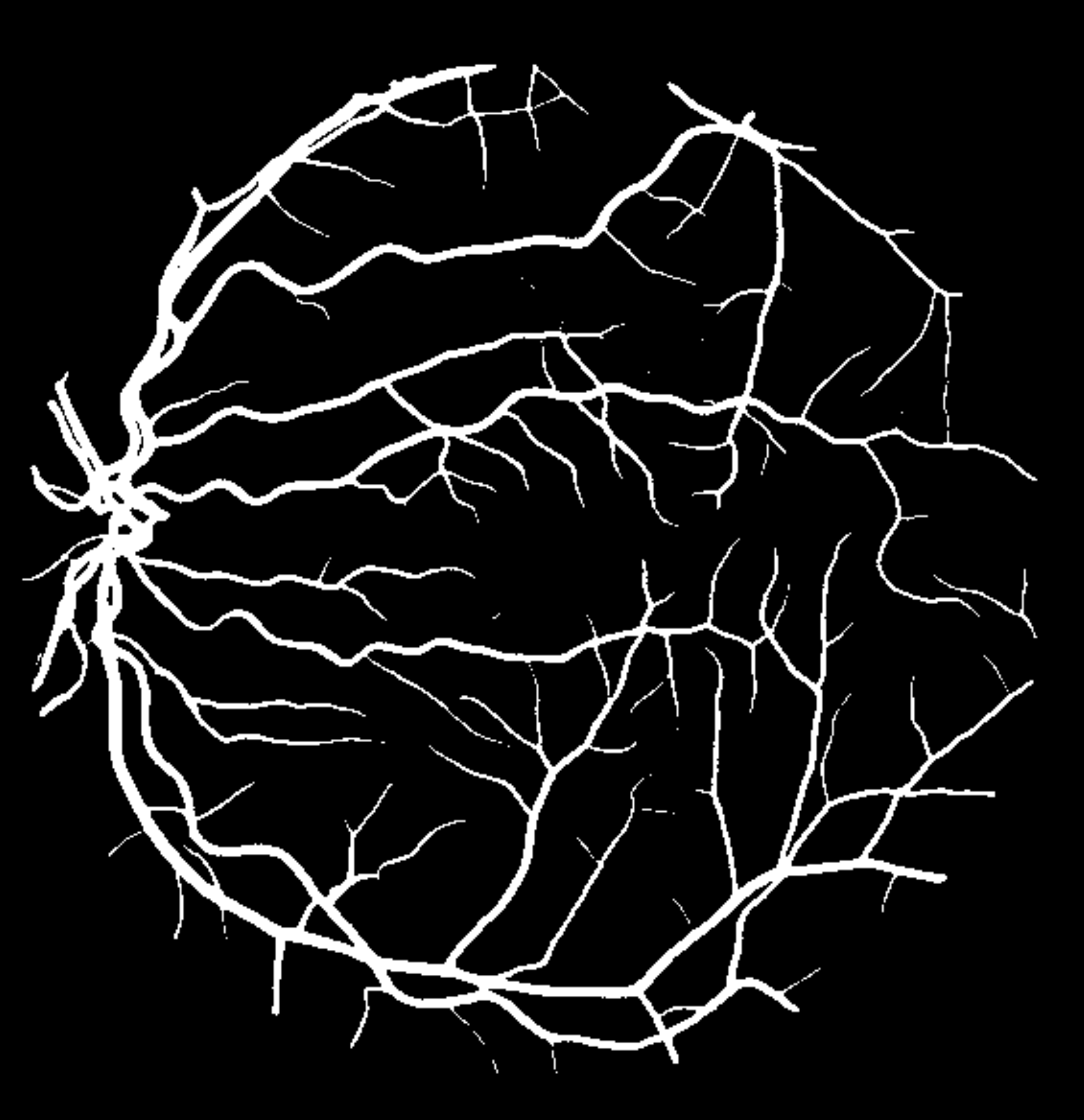}
}
\\
\subfigure{
\includegraphics[width=0.2\textwidth]{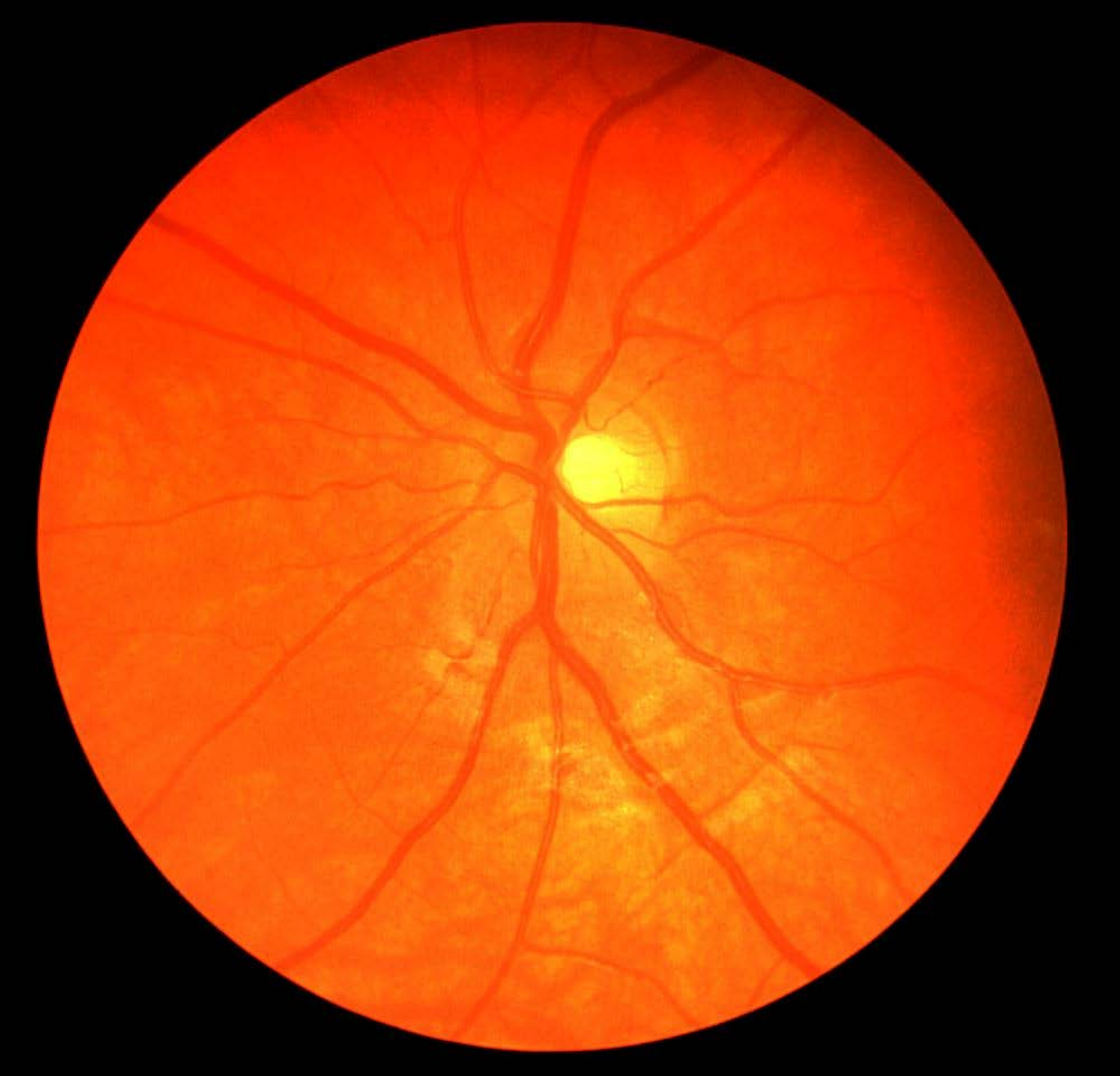}
}
\subfigure{
\includegraphics[width=0.2\textwidth]{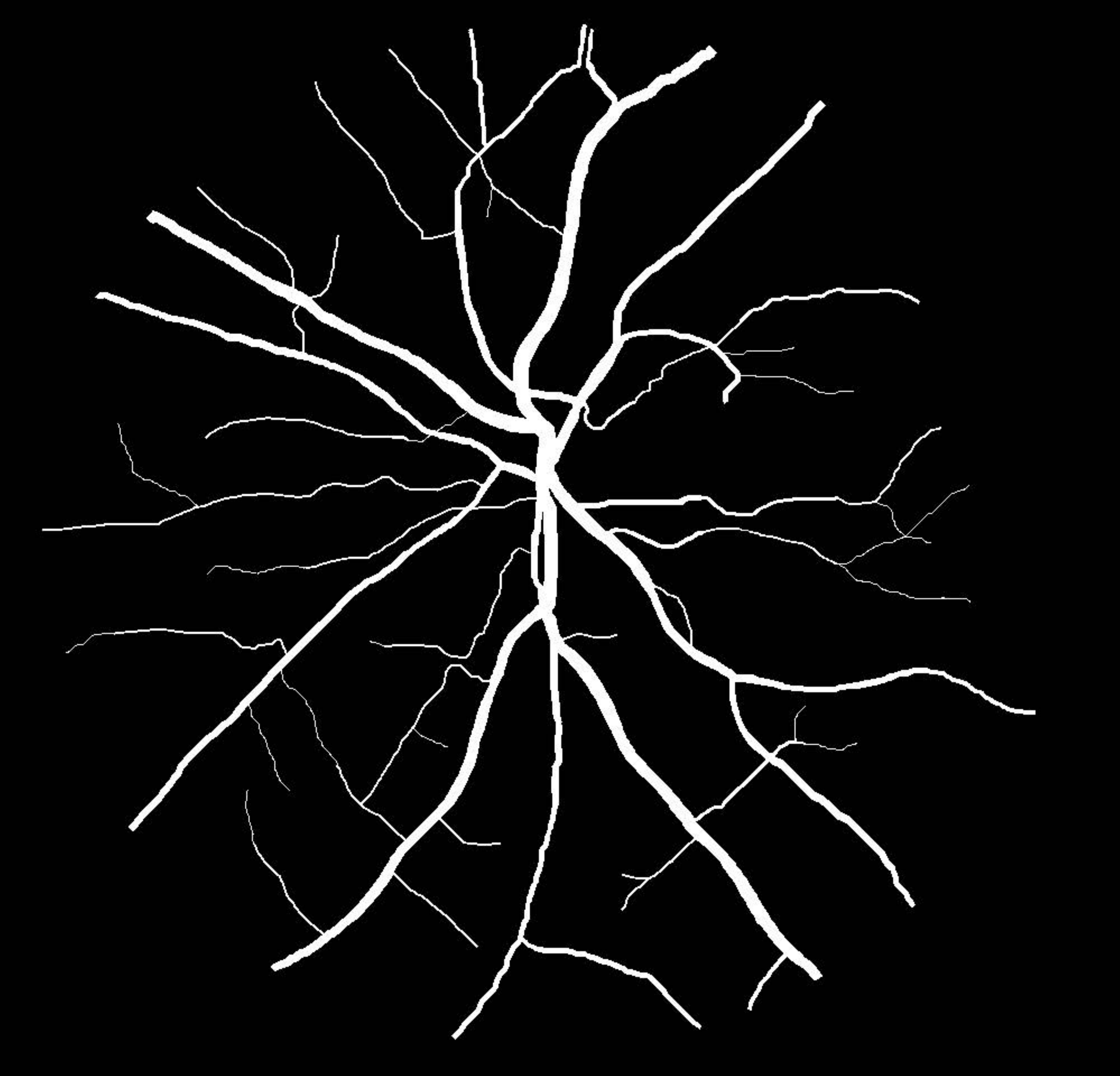}
}
\subfigure{
\includegraphics[width=0.2\textwidth]{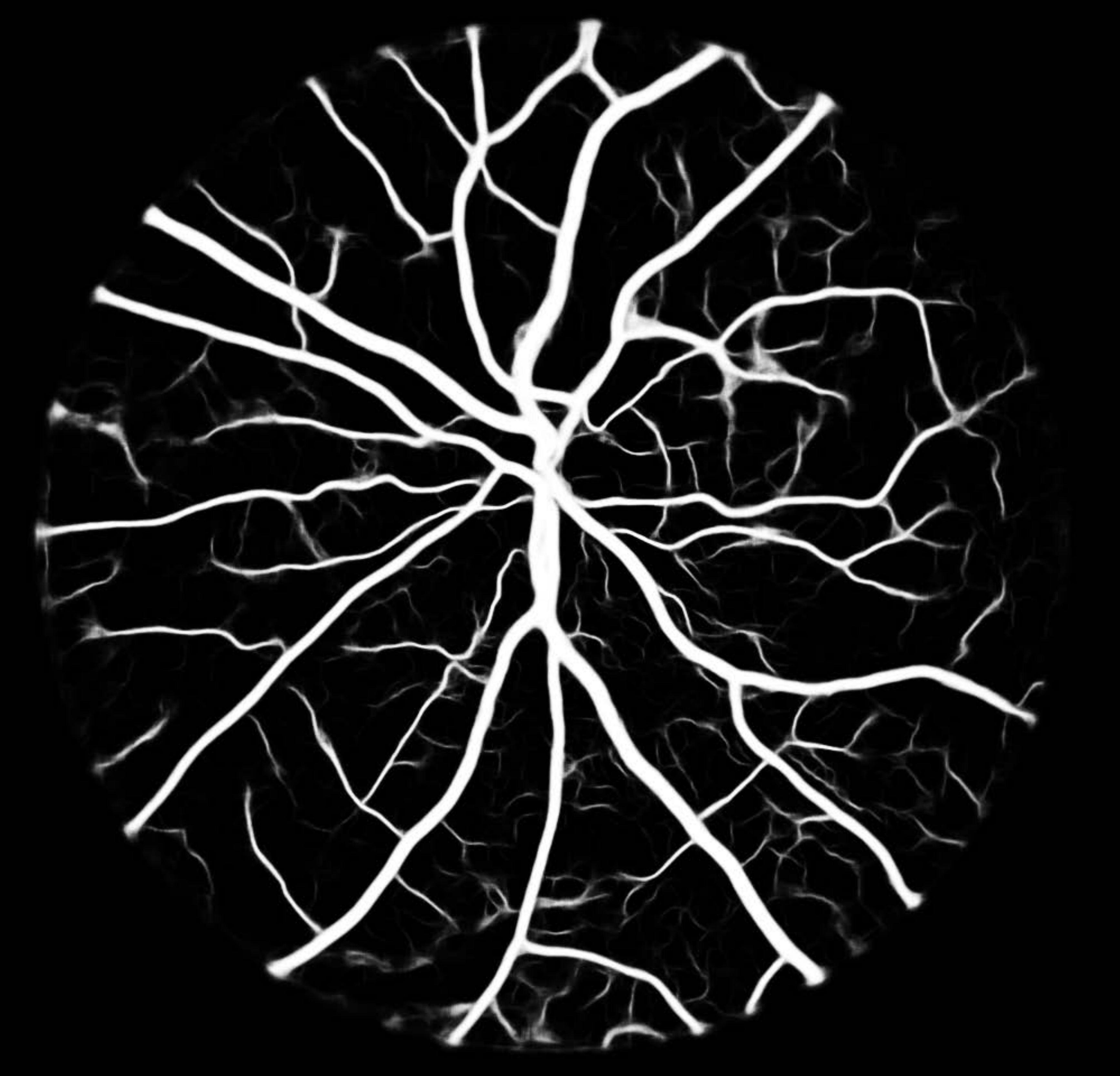}
}
\subfigure{
\includegraphics[width=0.2\textwidth]{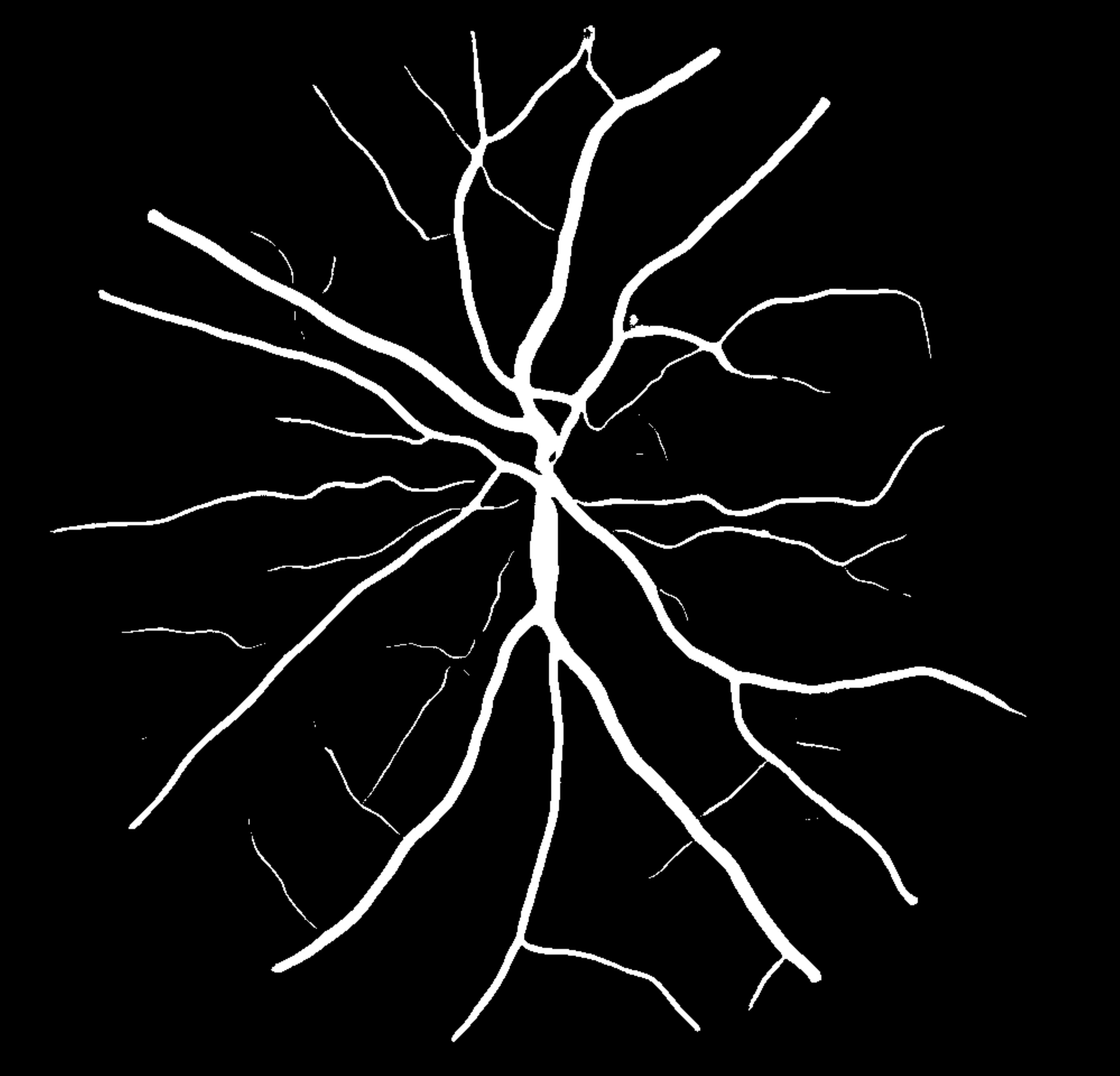}
}
\\
\subfigure{
\includegraphics[width=0.2\textwidth]{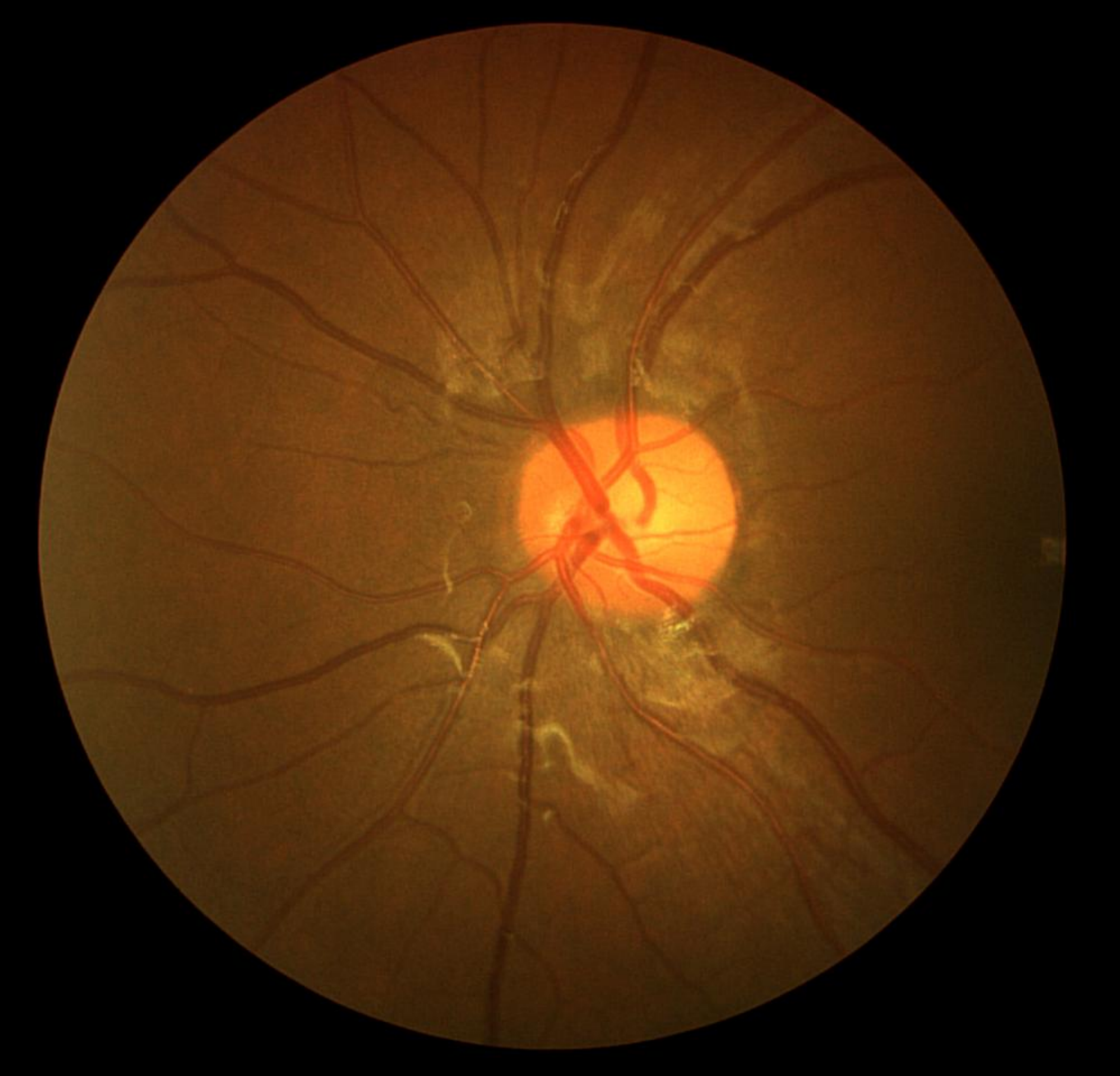}
}
\subfigure{
\includegraphics[width=0.2\textwidth]{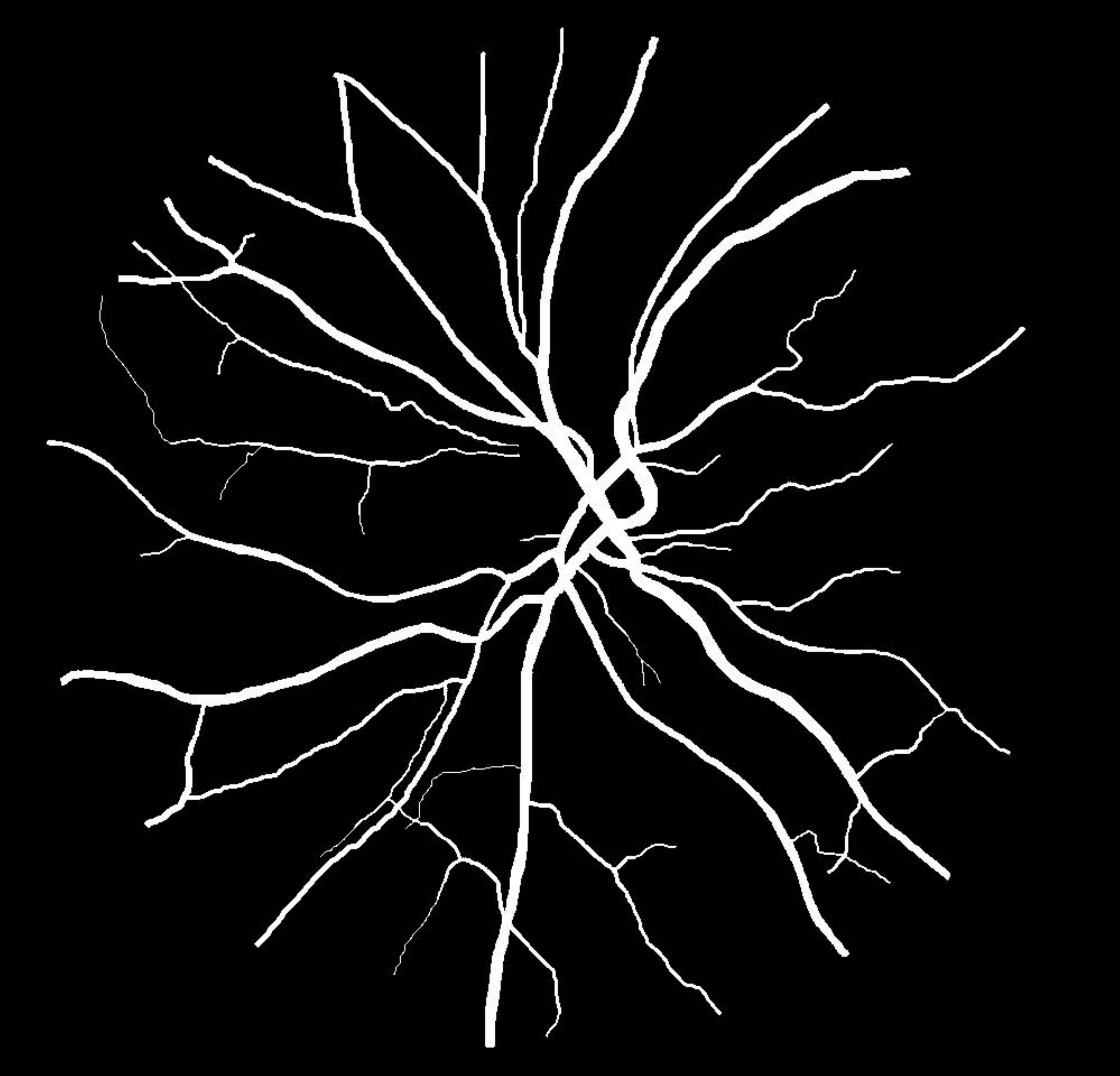}
}
\subfigure{
\includegraphics[width=0.2\textwidth]{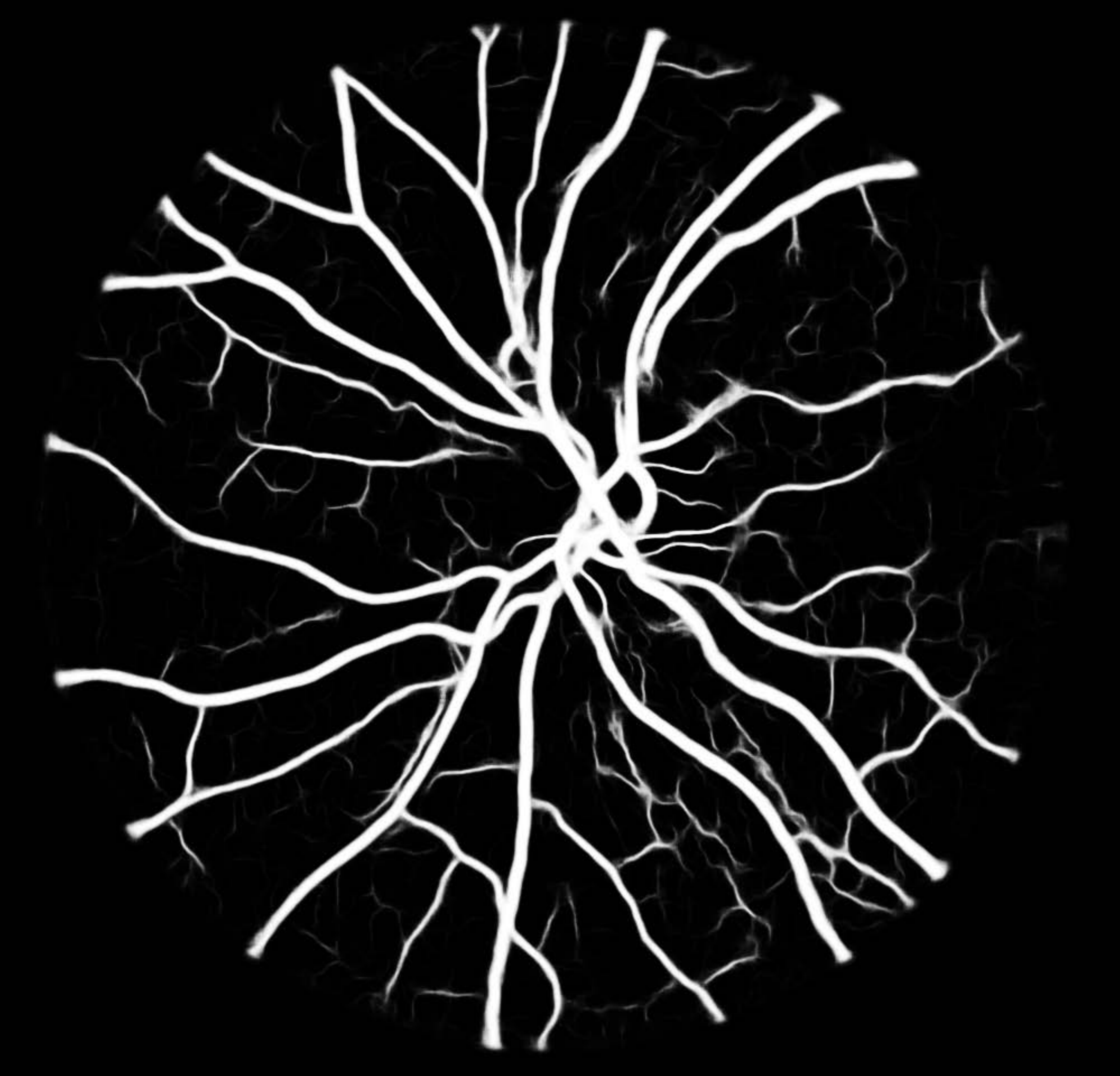}
}
\subfigure{
\includegraphics[width=0.2\textwidth]{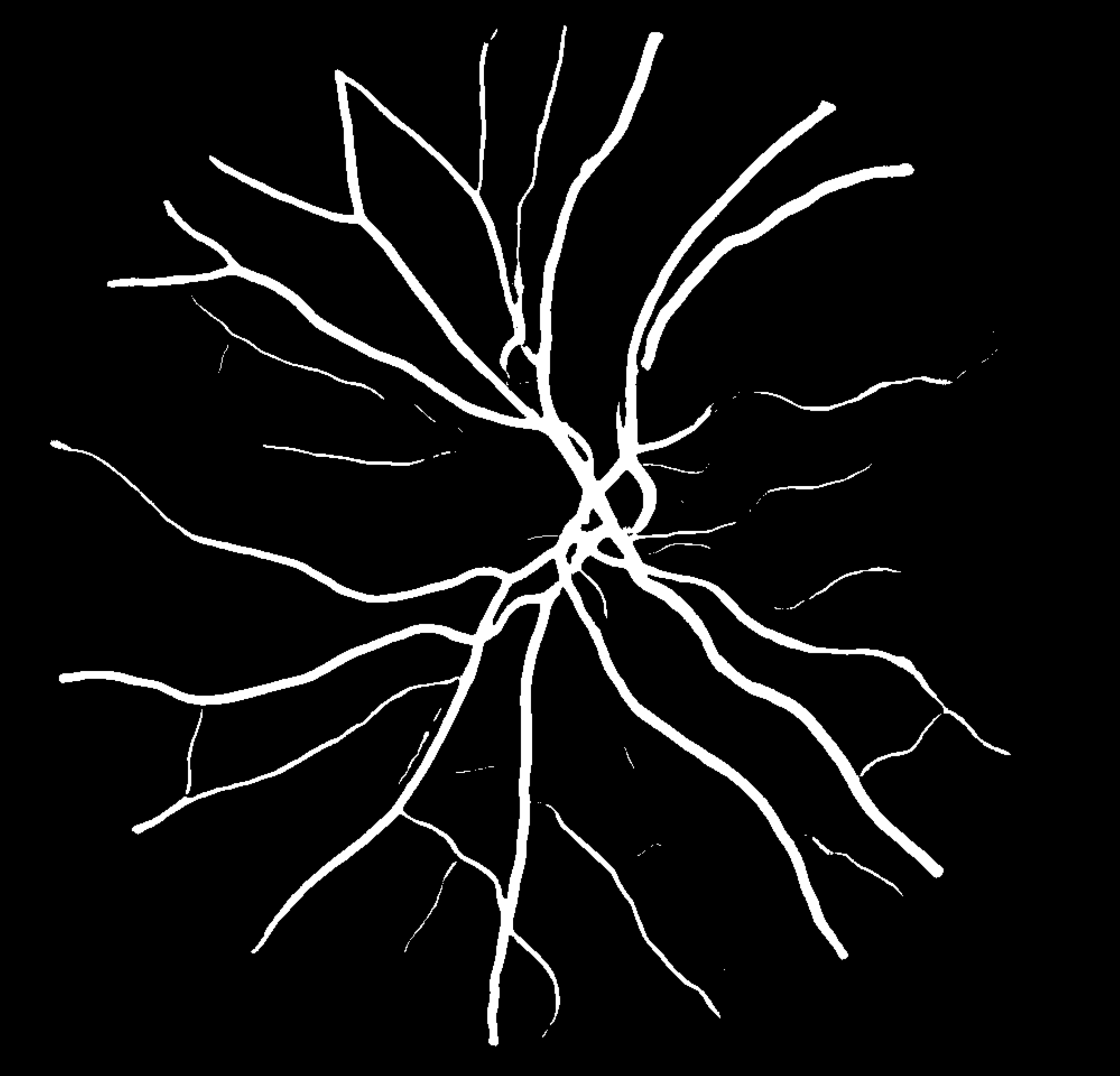}
}
\\
\subfigure{
\includegraphics[width=0.2\textwidth]{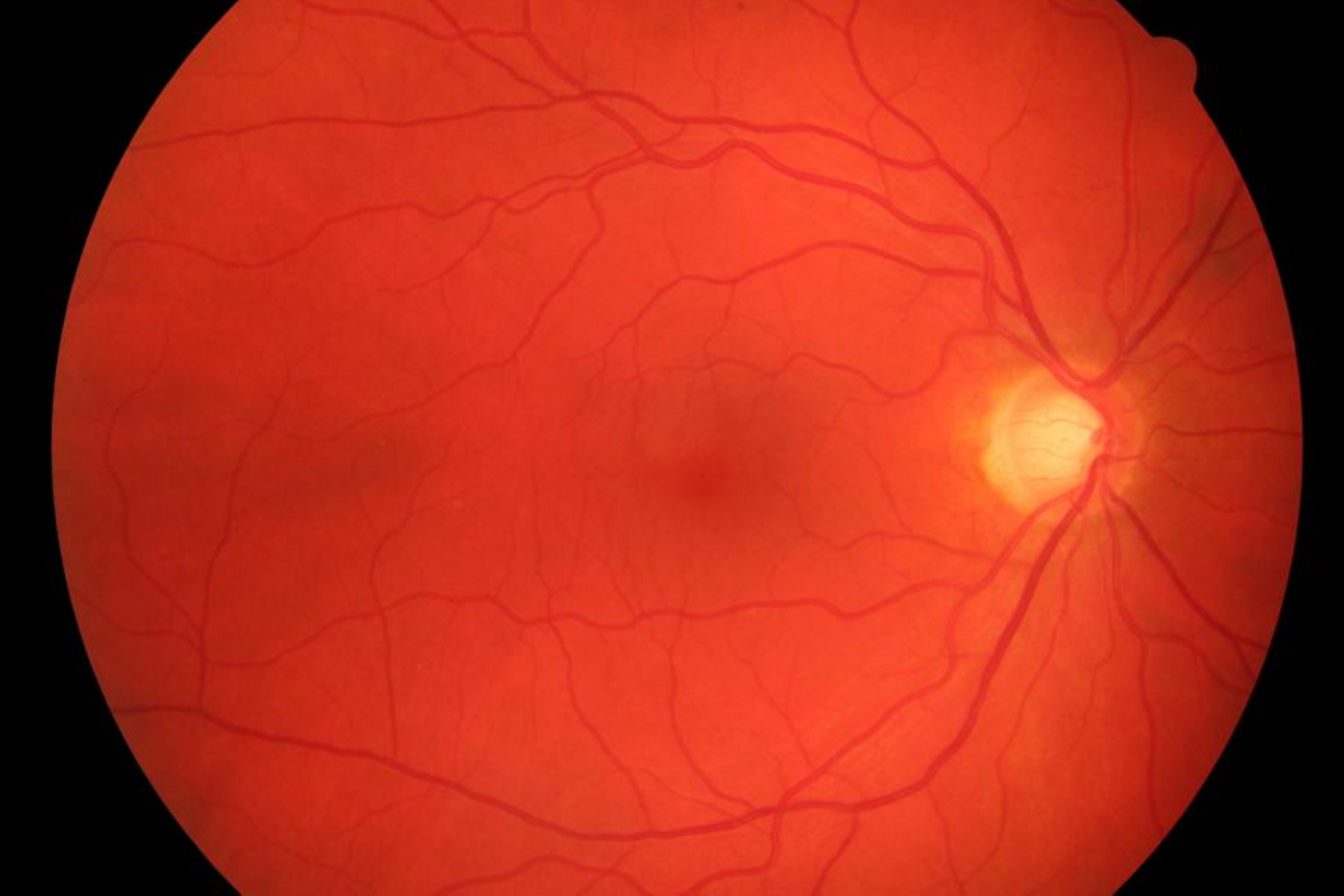}
}
\subfigure{
\includegraphics[width=0.2\textwidth]{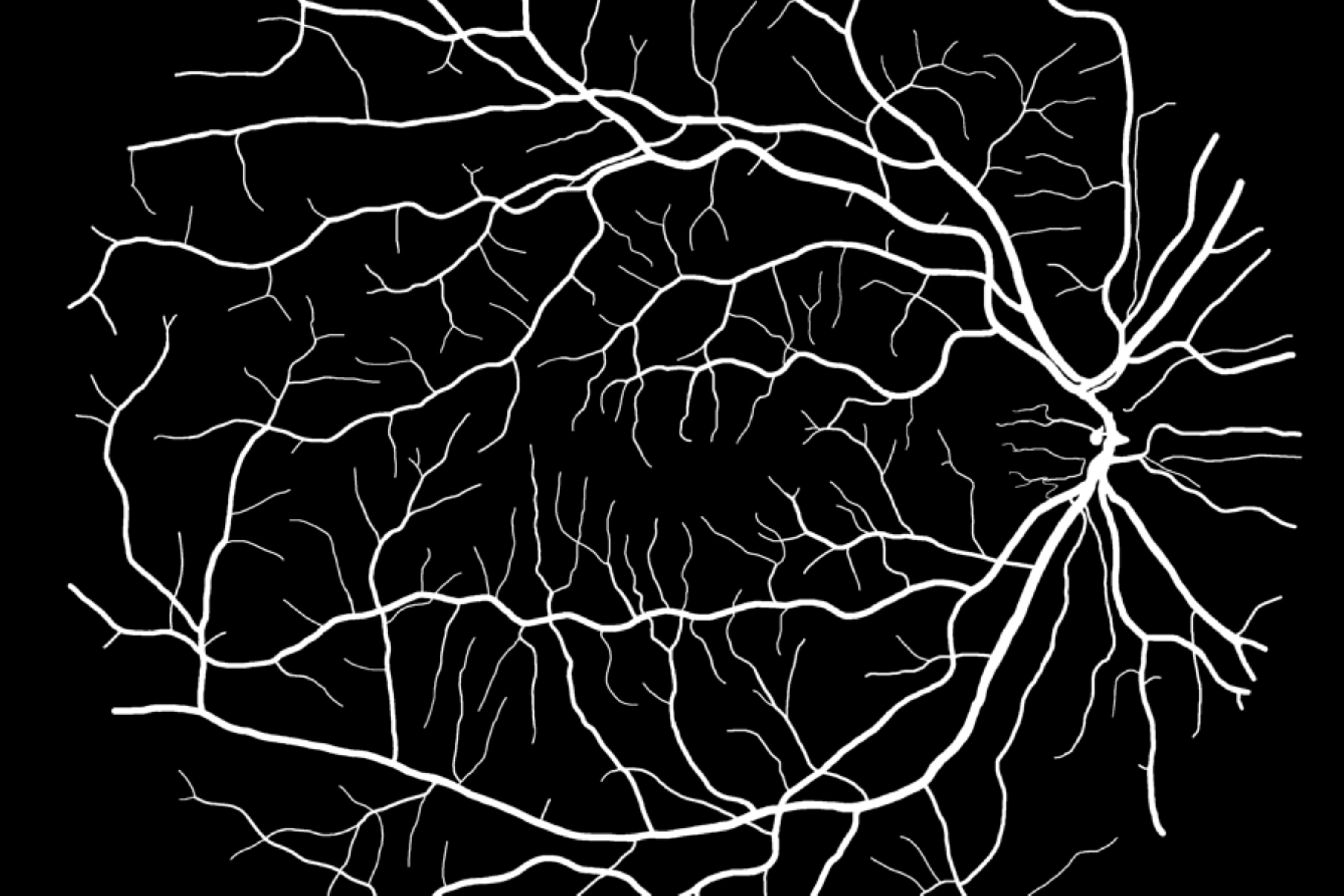}
}
\subfigure{
\includegraphics[width=0.2\textwidth]{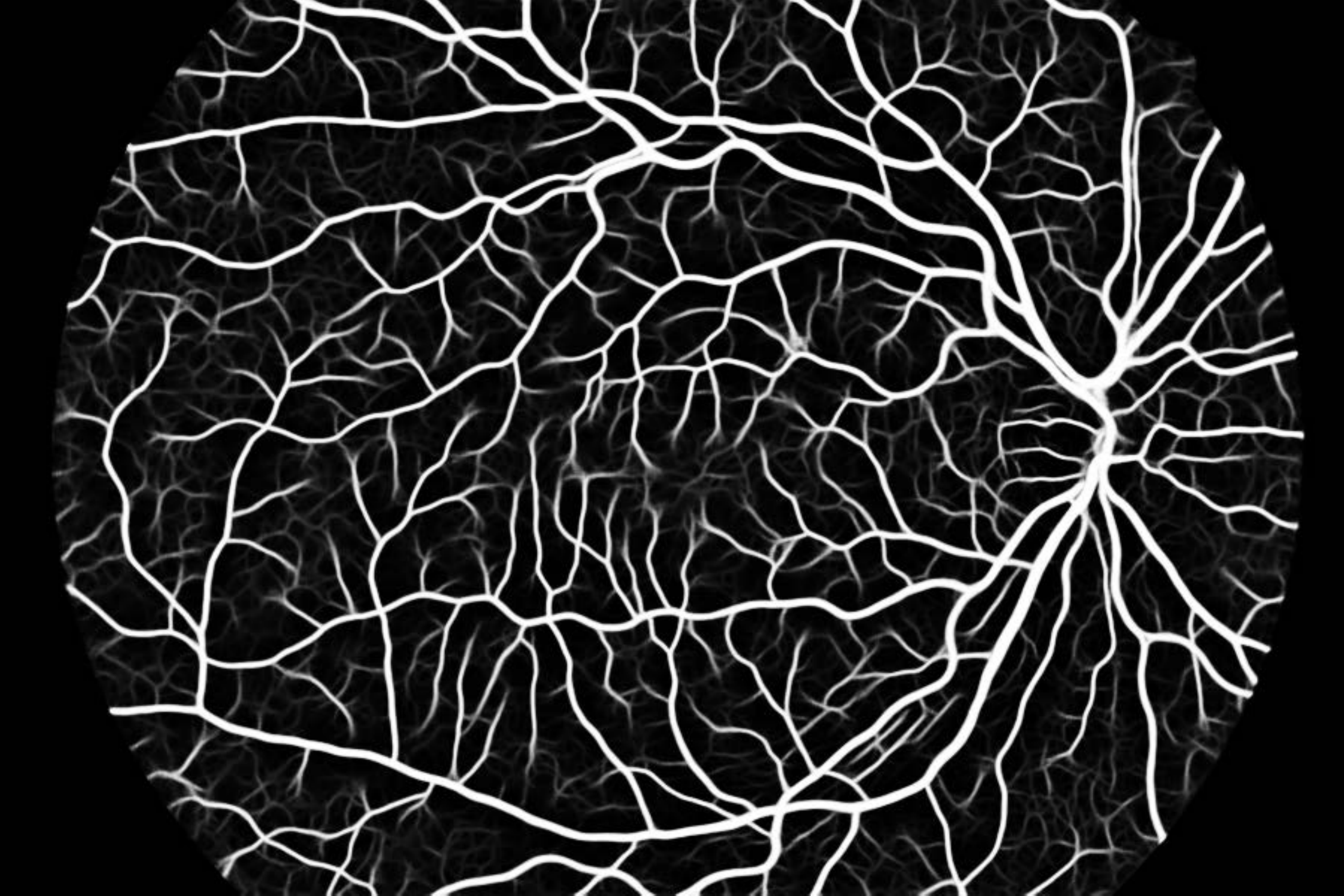}
}
\subfigure{
\includegraphics[width=0.2\textwidth]{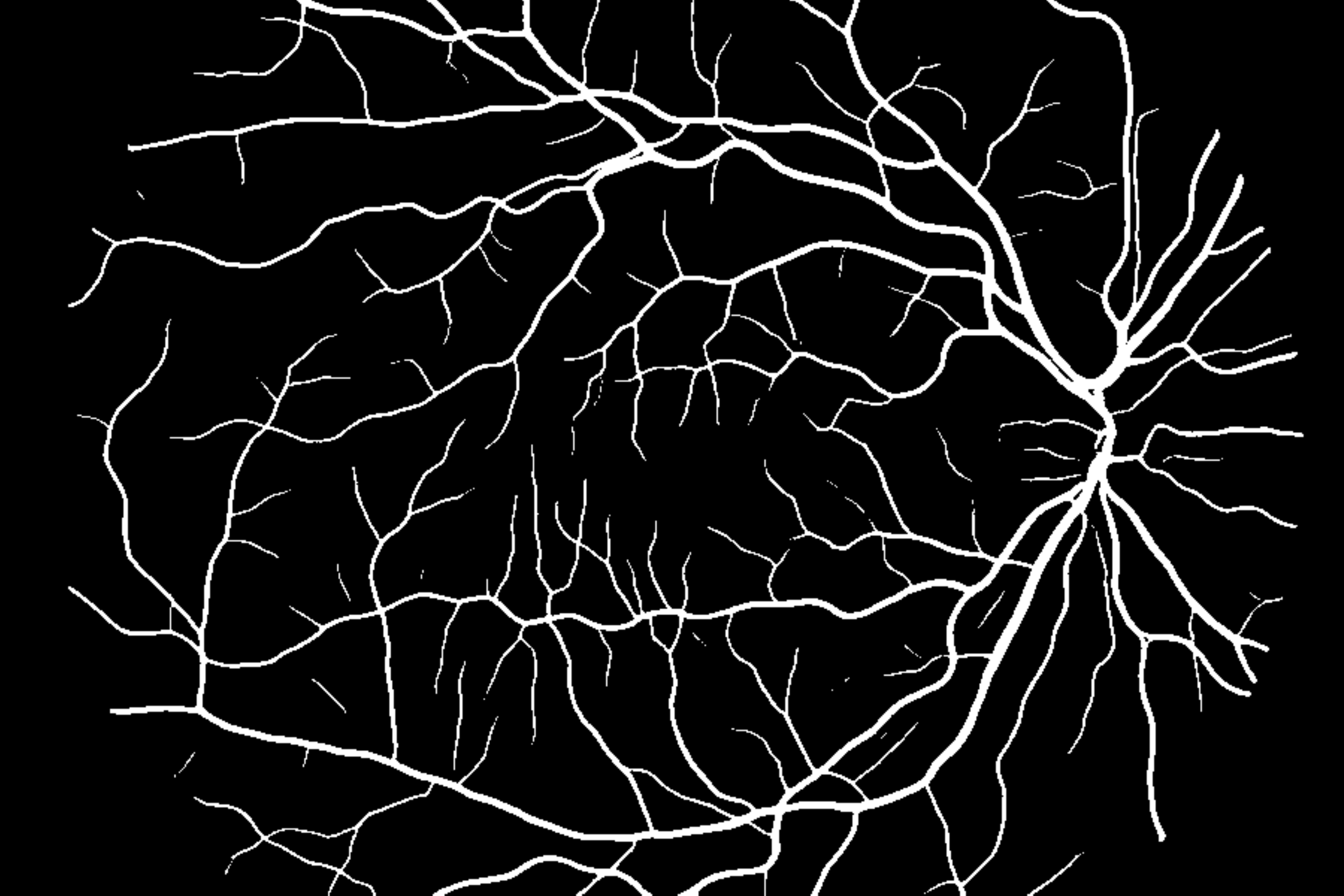}
}\\
\subfigure{
\includegraphics[width=0.2\textwidth]{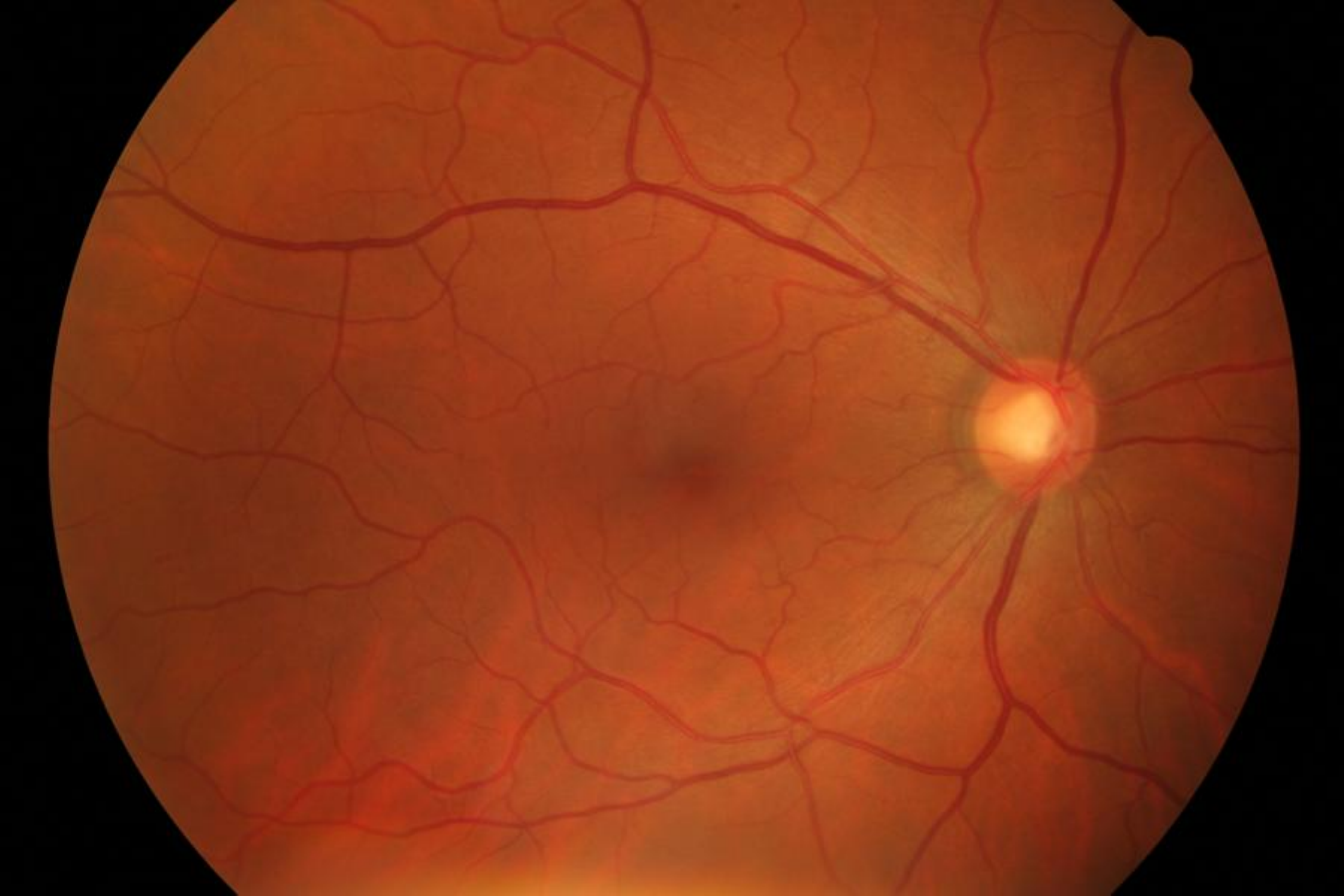}
}
\subfigure{
\includegraphics[width=0.2\textwidth]{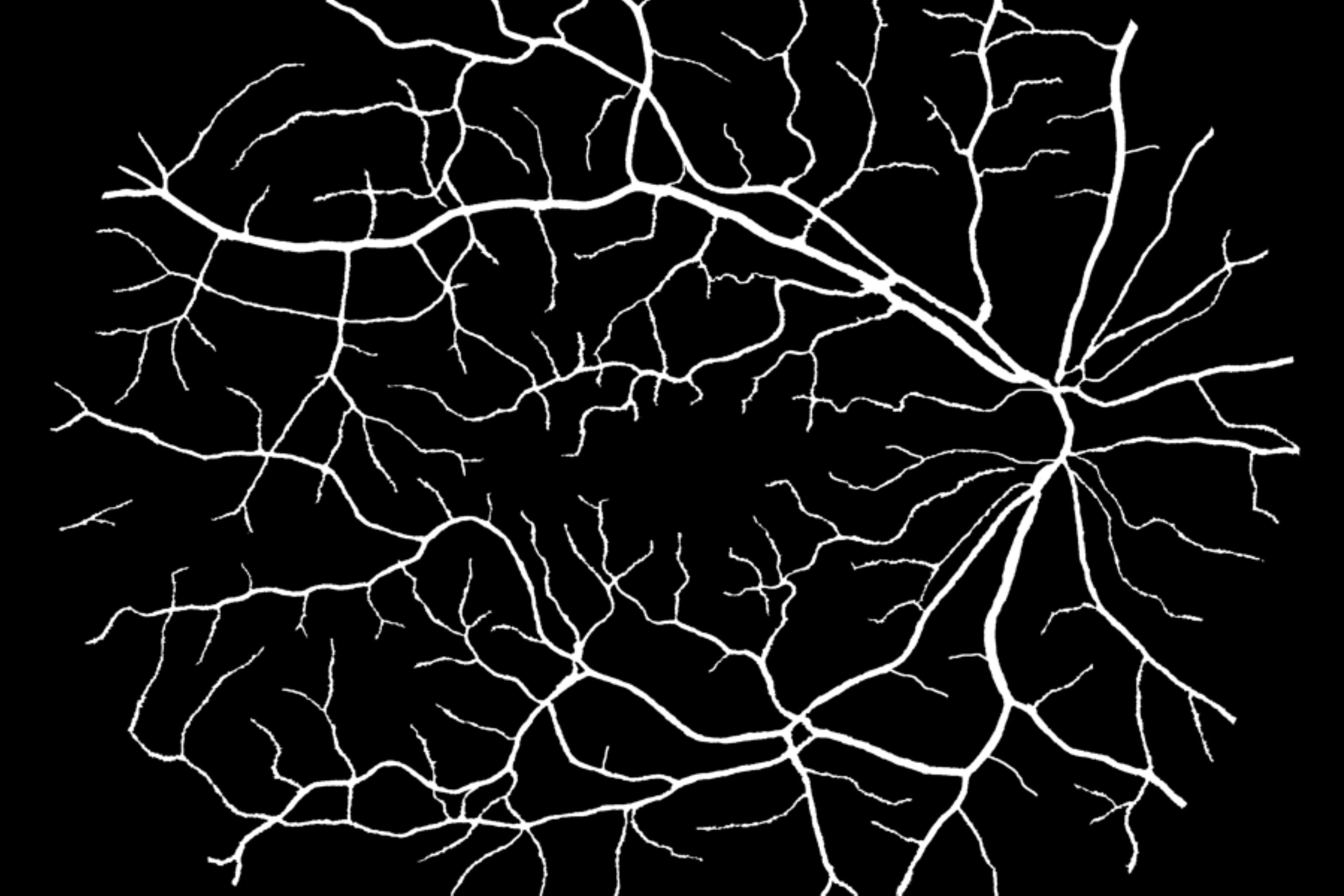}
}
\subfigure{
\includegraphics[width=0.2\textwidth]{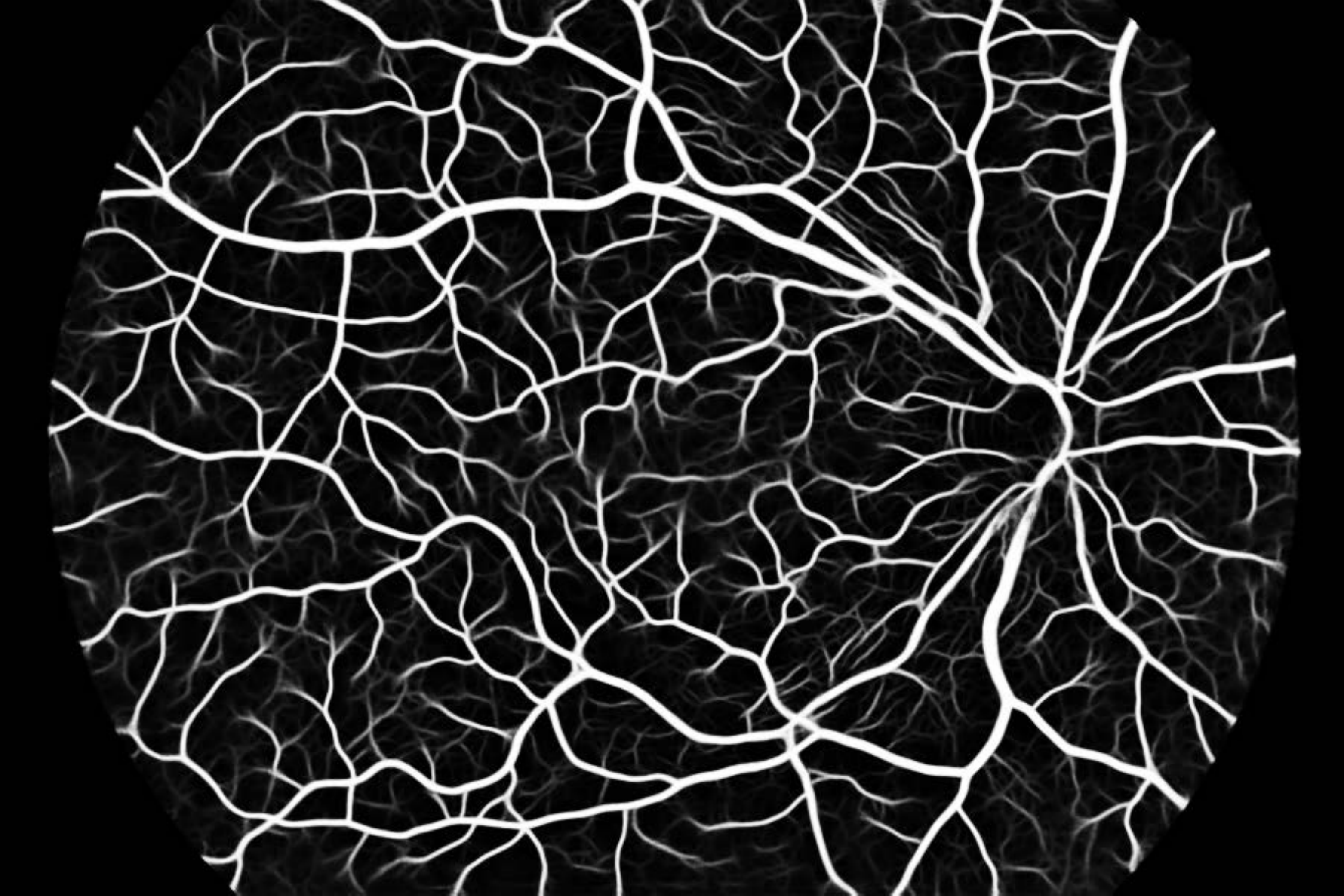}
}
\subfigure{
\includegraphics[width=0.2\textwidth]{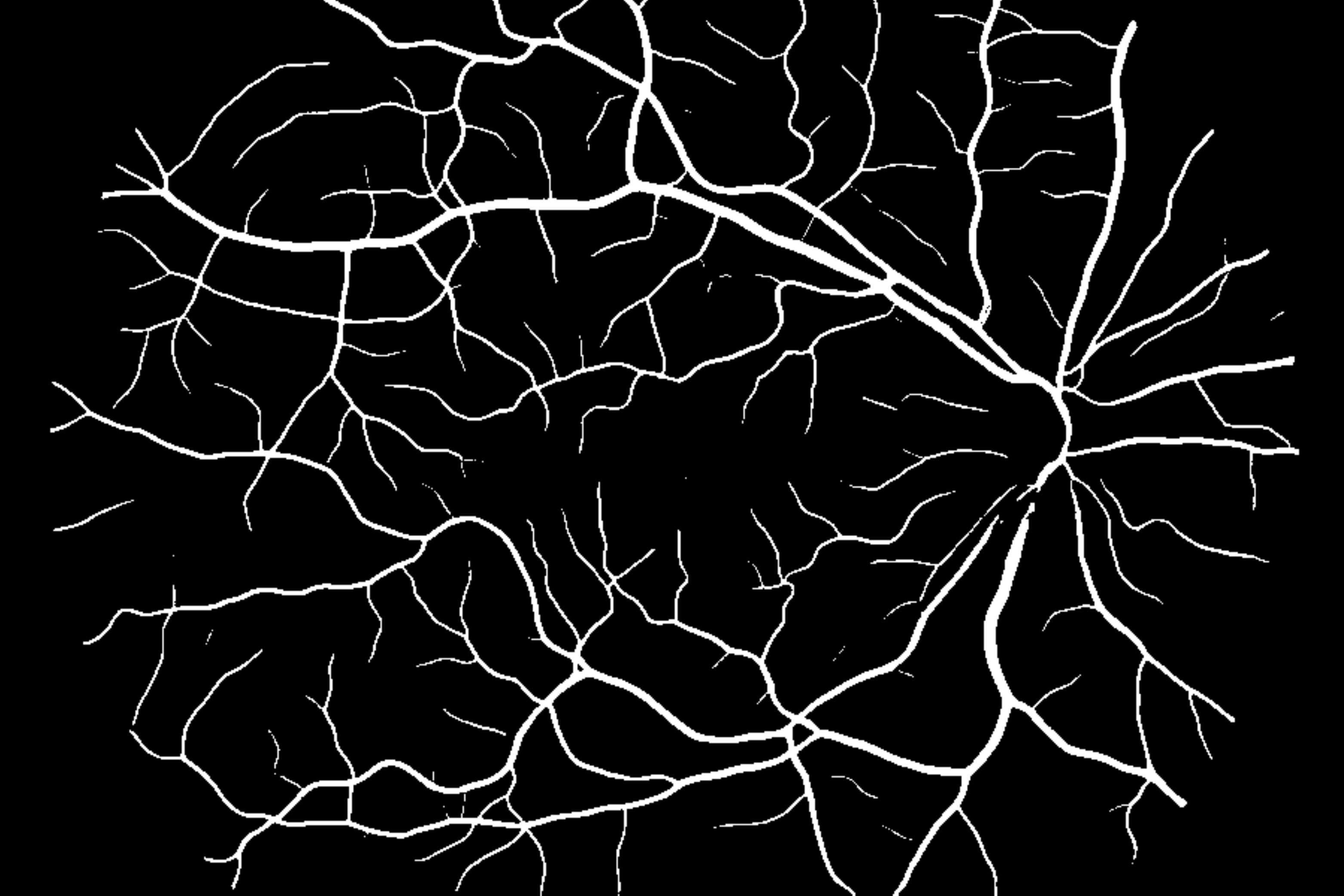}
}
\caption{Visualization of the segmentation maps.
From columns 1 to 4: fundus images, ground truth, probability maps, and binary maps.}
\label{fig:vessel_seg_maps}
\end{figure*}

To show the effectiveness of our proposed DPN, we present the segmentation probability maps and the corresponding binary maps in Fig.~\ref{fig:vessel_seg_maps}. We can observe that our model could detect both thin vessels and thick vessel trees, verifying the effectiveness of our proposed DP-Block.

\begin{figure}
\centering
\subfigure[Segmentation of extremely thin vessels]{
\includegraphics[width=0.28\textwidth]{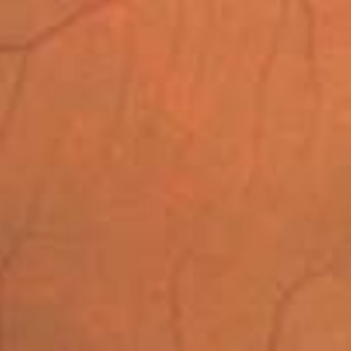}
\includegraphics[width=0.28\textwidth]{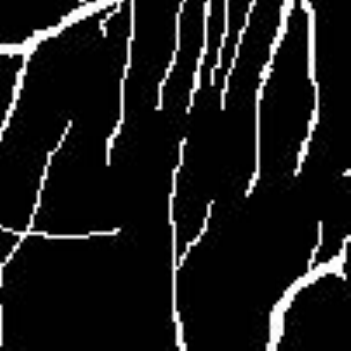}
\includegraphics[width=0.28\textwidth]{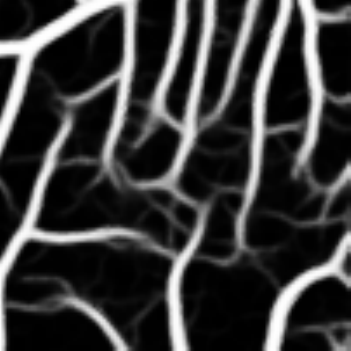}
}\\
\subfigure[Segmentation of low-contrast vessels]{
\includegraphics[width=0.28\textwidth]{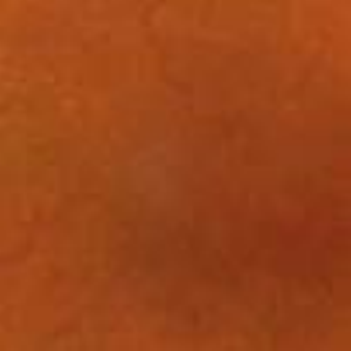}
\includegraphics[width=0.28\textwidth]{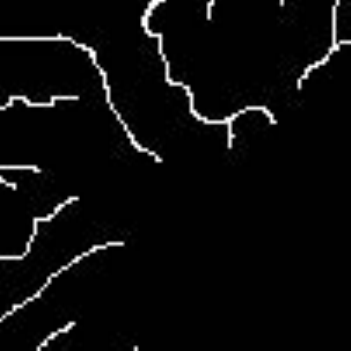}
\includegraphics[width=0.28\textwidth]{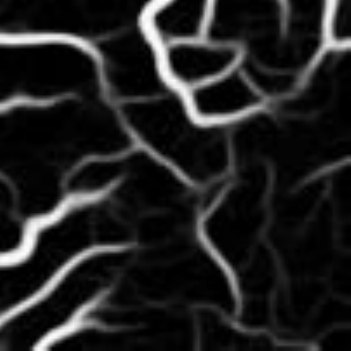}
}\\
\subfigure[Segmentation in the presence of hemorrhages]{
\includegraphics[width=0.28\textwidth]{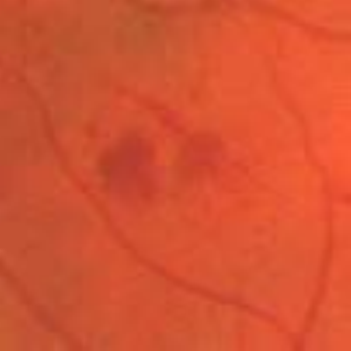}
\includegraphics[width=0.28\textwidth]{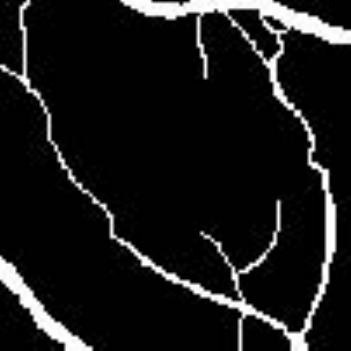}
\includegraphics[width=0.28\textwidth]{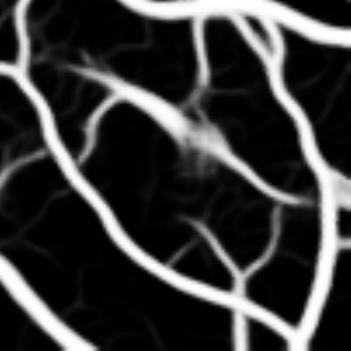}
}
\subfigure[Segmentation in the presence of microaneurysms]{
\includegraphics[width=0.28\textwidth]{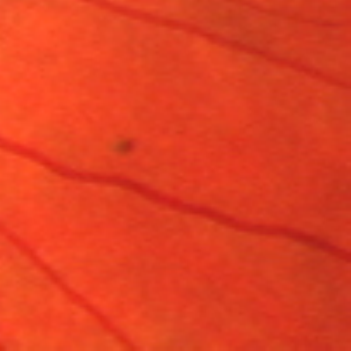}
\includegraphics[width=0.28\textwidth]{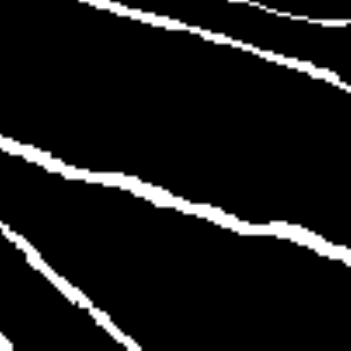}
\includegraphics[width=0.28\textwidth]{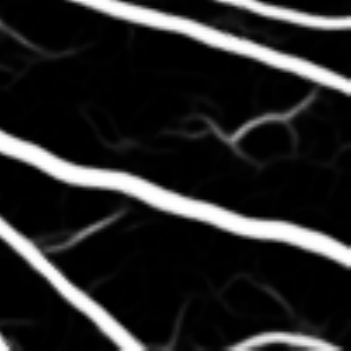}
}
\caption{Visualization of some challenging cases. From left to right: fundus images patches, ground-truth, and the segmentation probability maps generated by proposed DPN.}
\label{fig:vessel_seg_challenge}
\end{figure}

Moreover, we present several challenging cases in Fig.~\ref{fig:vessel_seg_challenge}.
We can observe that our model could detect thin vessels with only one-pixel width, as DPN always preserves the spatial information.
In addition, our model is able to segment some extremely thin vessels with low contrast near the macula.
In the third row of Fig.~\ref{fig:vessel_seg_challenge}, there exist two lumps of hemorrhage, which shares similar local features with vessels. As the DPN could capture structural information, as a result, the DPN is robust to the presence of hemorrhages. Also, for some true vessels not annotated, our model could segment well. In summary, the proposed method could segment thick and thin vessels, and robust to noise.

\subsection{Ablation Study}
\subsubsection{Effectiveness of Auxiliary Losses}
In order to verify the impact of auxiliary losses on the final segmentation performance of the model, we removed all three auxiliary losses in DPN and the model was trained again under the same settings.
The experimental results were summarized in Table~\ref{table:comp_loss}. We can observe that almost all evaluation metrics were decreased after removing auxiliary losses.
Specifically, the F1-score was decreased over 0.3\% on all three datasets.
This part of the experiments verifies the rationality and effectiveness of adopting auxiliary losses in DPN.

\begin{table}
\setlength{\tabcolsep}{2pt}
\begin{center}
\caption{Comparison results of employing auxiliary losses or not (best results shown in bold).}
\begin{tabular}{ccccccc}
\hline
Dataset &Auxiliary Loss? &Se &Sp &Acc &AUC &F1\\
\hline
\multirow{2}{*}{DRIVE}
&No       &0.7874 &\textbf{0.9810}  &0.9564 &0.9808 &0.8259\\
&Yes      &\textbf{0.7934} &\textbf{0.9810}  &\textbf{0.9571} &\textbf{0.9816} &\textbf{0.8289}\\
\hline
\multirow{2}{*}{CHASE\_DB1}
&No    &0.7805 &0.9833 &0.9649 &0.9840 &0.8058 \\
&Yes   &\textbf{0.7839} &\textbf{0.9842} &\textbf{0.9660} &\textbf{0.9860} &\textbf{0.8124}\\
\hline
\multirow{2}{*}{HRF}
&No   &0.7887 &0.9759 &0.9583 &0.9679 &0.7792\\
&Yes  &\textbf{0.7926} &\textbf{0.9764} &\textbf{0.9591} &\textbf{0.9697} &\textbf{0.7835}\\
\hline
\end{tabular}
\label{table:comp_loss}
\end{center}
\end{table}

\subsubsection{Effectiveness of Multi-scale Feature Fusion in DP-Block}
In order to verify the effectiveness of multi-scale feature fusion in DP-Block, we conduct three groups of experiments, i.e., the DP-Block contains only the first branch, the DP-Block contains the first and the second branch, and the DP-Block contains all of three branches.
Experimental results are summarized in Table~\ref{table:comp_pool}.
We can observe that, after adding the second and the third branch, the segmentation performance has been improved over DRIVE and CHASE\_DB1 datasets. To be specific, the F1-score of DPN with three branches is over 1.09\% higher than that of DPN with only the first branch. This part of the experiments reveals that it is necessary to adopt multi-scale feature fusion within DP-Block.

\begin{table}
\setlength{\tabcolsep}{2pt}
\begin{center}
\caption{Comparison results of multi-scale feature fusion in DP-Block (OS is short for output stride, and OS1/OS2/OS4 corresponds to the first/second/third branch of DP-Block, respectively)}
\begin{tabular}{clccccccc}
\hline
Dataset &Feature Fusion &Se &Sp &Acc &AUC &F1 &SSIM &PSNR\\
\hline
\multirow{3}{*}{DRIVE}
&OS1      &0.7875 &0.9813  &0.9566 &0.9808 &0.8262 &0.5303 &\textbf{13.8216}\\
&+OS2     &0.7874 &\textbf{0.9816}  &0.9569 &0.9810 &0.8277 &0.5093 &13.5204 \\
&+OS2+OS4 &\textbf{0.7934} &0.9810  &\textbf{0.9571} &\textbf{0.9816} &\textbf{0.8289} &\textbf{0.5500} &13.7672\\
\hline
\multirow{3}{*}{CHASE\_DB1}
&OS1        &0.7631 &\textbf{0.9843} &0.9642 &0.9826 &0.8015 &0.5507 &14.1293\\
&+OS2       &0.7684 &0.9841 &0.9645 &0.9838 &0.8035 &0.5379 &13.4846\\
&+OS2+OS4   &\textbf{0.7839}&0.9842 &\textbf{0.9660} &\textbf{0.9860} &\textbf{0.8124} &\textbf{0.6602} &\textbf{14.4583}\\
\hline
\end{tabular}
\label{table:comp_pool}
\end{center}
\end{table}

\subsubsection{The Number of Convolutional Filters}
Table~\ref{table:comp_num_filters} demonstrates the ablation results of various number of convolutional filters. First, setting C0, C1, and C2 to 16, 8, 8 for each DP-Block outperforms other configurations over DRIVE and CHASE\_DB1 datasets. Second, improving segmentation speed, that is, such thin convolutional filters can significantly accelerate inference speed.
Third, thick convolutional filters mean high memory consumption and high optimization difficulty.
At last, considering the trade-off between segmentation accuracy and inference speed, we set C0, C1, and C2 to 16, 8, and 8 in DP-Block.

\begin{table}
\setlength{\tabcolsep}{2pt}
\begin{center}
\caption{Ablation studies on the number of convolutional filters.}
\begin{tabular}{ccccccccccc}
\hline
Dataset &C0 &C1 &C2 &Se &Sp &Acc &AUC &F1 &SSIM &PSNR\\
\hline
\multirow{4}{*}{DRIVE}
&8 &4 &4   &0.7894 &0.9802 &0.9559 &0.9806 &0.8242 &0.5150 &13.7664\\
&8 &8 &8   &0.7840 &\textbf{0.9815} &0.9563 &0.9808 &0.8254 &0.5141 &13.7618\\
&16 &8 &8  &0.7934 &0.9810  &\textbf{0.9571} &\textbf{0.9816} &\textbf{0.8289} &\textbf{0.5500} &\textbf{13.7672}\\
&24 &12 &12&\textbf{0.7942} &0.9808 &\textbf{0.9571} &0.9813 &0.8283  &0.5023 &13.5176\\
\hline
\multirow{4}{*}{CHASE\_DB1}
&8 &4 &4    &0.7563 &\textbf{0.9846} &0.9639 &0.9833 &0.7998 &0.5620 &13.8856\\
&8 &8 &8    &0.7485 &0.9845 &0.9631 &0.9820 &0.7950 &0.5703 &14.3965\\
&16 &8 &8   &\textbf{0.7839}&0.9842 &\textbf{0.9660} &\textbf{0.9860} &\textbf{0.8124} &\textbf{0.6602} &\textbf{14.4583}\\
&24 &12 &12 &0.7747 &0.9835 &0.9645 &0.9838 &0.8045 &0.5215 &13.5914 \\
\hline
\end{tabular}
\label{table:comp_num_filters}
\end{center}
\end{table}

\section{Conclusion}
\label{sec:con}
Deep learning models have been applied to retinal vessel segmentation in recent years, and achieve remarkable performance.
In this paper, we propose a deep model, called DPN to segment retinal vessel trees. Different from U-Net and FCN, in which detailed spatial information is sacrificed to learn structured information. Our method could preserve the detailed information all the time via maintaining a high resolution throughout the whole process, benefiting locating the vessel boundaries accurately.
To accomplish this goal, we present the DP-Block further, where multi-scale fusion is adopted to preserve both detailed information and learn structural information. In order to show the effectiveness of our method, DPN is trained from scratch over three publicly available datasets: DRIVE, CHASE\_DB1, and HRF.
Experimental results show that our method shows competitive/better performance in terms of F1-score and segmentation speed with only about 120k parameters. Specifically, the segmentation speed of our method is over 20-160$\times$ faster than other state-of-the-art methods on the DRIVE dataset.
In summary, considering the segmentation accuracy, segmentation speed, and model size together, our model shows superior performance and is suitable for real-world application.
Meantime, there are some drawbacks of DPN. For example, it cannot process high-resolution fundus images directly, and there are discontinuous vessel patches in the binary map.
In the future, we aim to extend our method and develop robust deep models for fundus microaneurysms segmentation.
At last, the source code of our method is available at \url{https://github.com/guomugong/DPN}.

\section*{Funding}
This work is supported by PhD research startup foundation of Xi'an University of Architecture and Technology (No.1960320048).

\section*{Conflict of interest}
The authors have no conflicts of interest to declare that are relevant to the content of this article.
\section*{Availability of data and material}
All databases utilized in this publication are publicly available, namely DRIVE~\cite{drive}, CHASE\_DB1~\cite{chase}, and HRF~\cite{budai2013robust}.

\section*{Code availability}
The source code will be available at \url{https://github.com/guomugong/DPN}.


%\begin{thebibliography}{}

\bibliographystyle{spbasic}
\bibliography{dpn}
\end{document}